\documentstyle[11pt]{article}

\def\NPB#1#2#3{Nucl. Phys. {\bf B#1} (#2) #3}
\def\PLB#1#2#3{Phys. Lett. {\bf B#1} (#2) #3}

\def\PRC#1#2#3{Phys. Rev. {\bf C#1} (#2) #3}
\def\PRD#1#2#3{Phys. Rev. {\bf D#1} (#2) #3}

\def\PRL#1#2#3{Phys. Rev. Lett. {\bf #1} (#2) #3}

\def\HPH#1{hep--ph{}/{}#1}

\def\HEX#1{hep--ex{}/{}#1}

\def\etal{\hbox{\it et al.}{}}

\def\nn{\hspace{2mm}}
\def\sss{\scriptscriptstyle}

\newcommand{\GeV}{\mbox{\rm GeV}}

\def\sleq{\raisebox{-.6ex}{${\textstyle\stackrel{<}{\sim}}$}}
\def\sgeq{\raisebox{-.6ex}{${\textstyle\stackrel{>}{\sim}}$}}
\def\Tr{\rm{Tr}}

\def\Bar#1{\overline{#1}}

\def\ket#1{\left| #1\right\rangle}

\def\sVEV#1{\left\langle #1\right\rangle}

\def\cL{{\cal L}}

\begin{document}
\begin{titlepage}
\hfill
\vbox{
    \halign{#\hfil        \cr
           CERN-TH/2000-119 \cr
           NBI-HE-00-18    \cr
           hep-ph/0004137 \cr
           } 
      }  
\vspace*{20mm}
\begin{center}
{\Large {\bf  Neutrino mass matrix suppression by Abelian charges with see-saw mechanism}\\}

\vspace*{15mm}
\vspace*{1mm}
{\ H. B. Nielsen}$^{\rm a,b,}$\footnote[3]{E-mail: holger.bech.nielsen@cern.ch; hbech@nbi.dk}
and {\ Y. Takanishi}$^{\rm b,}$\footnote[4]{E-mail: yasutaka@nbi.dk}

\vspace*{1cm} 
{\it $^a$Theory Division, CERN, \\
CH-1211 Geneva 23, Switzerland}\\
\vskip .5cm
{\it $^b$The Niels Bohr Institute,\\
Blegdamsvej 17, DK-2100 Copenhagen {\O}, Denmark}\\

\vspace*{.5cm}
\end{center}

\begin{abstract}
We have investigated a neutrino mass matrix model without 
supersymmetry including three see-saw right-handed neutrinos 
around order $10^{12}~\GeV$ masses, aiming at a picture with 
all small numbers explained as being due to approximately 
conserved gauge charges. The prediction of the solar 
neutrino mixing angle is given by 
$\sin^22\theta_{\odot}= 3\,{ +3\atop -2 }\times10^{-2}$; 
in fact, the solar mixing angle is, apart from detailed order 
unity corrections, equal to the Cabibbo angle. Furthermore the 
ratio of the solar neutrino mass square difference to that for 
the atmospheric neutrino oscillation is predicted 
to $6\,{ +11\atop -4 }\times10^{-4}$ and is given by the 
same Cabibbo angle related parameter $\xi$ as $6\,\xi^4$.
\end{abstract}
\vskip 4cm

April 2000
\end{titlepage}

\newpage

\section{Introduction}
\subsection{Neutrino data}
\indent

According to several experiments \cite{SK,Kamioka,talks,Macro,Chooz} 
it now seems that neutrino oscillations have to be taken seriously, and 
in all likelihood we should interpret such neutrino oscillations as 
being due to neutrino masses. We consider it relatively hopeless to 
incorporate the neutrino oscillations observed by LSDN \cite{LSND} which 
are partly excluded by Karmen 2 \cite{Karmen2}, especially insofar as 
in the present article we adopt the philosophy that only 
particles which are mass protected by the Standard Model (SM) gauge 
symmetries are sufficiently light to be observed at present. 
Thus we do not have sterile neutrinos in the model, which we will explain.

The numbers characterising the positively observed neutrino oscillations 
are concentrated in the fits of the atmospheric neutrino oscillations 
from Super-Kamiokande and Kamiokande (and neglecting the LSDN) 
\cite{SK,Kamioka}, leading to the presumed mixing 
of the $\nu_{\mu}$ with the $\nu_{\tau}$ corresponding to a mass 
square difference between the two mainly involved eigenmass neutrinos 
of
\begin{equation}
\Delta m_{\rm atm}^{2} \approx (2-6) 
\times 10^{-3}\;{\rm eV^{2}} 
\end{equation}%
\noindent
and a mixing angle
\begin{equation}
\sin^{2}2\theta_{\rm atm} \geq 0.82 \nn.\label{atmosangl} 
\end{equation}
\indent
On the other hand, the solar neutrino problems can be solved through
either vacuum or matter-enhanced Mikhyev-Smirnov-Wolfenstein (MSW) 
oscillation \cite{MSW}. The combination of the various solar neutrino 
experiments \cite{solar} allows - if taken seriously -
three different regions of fitting of mass square difference and mixing angle:

\noindent
(i) a small mixing angle (SMA): 
\begin{eqnarray}
\Delta m_{\odot }^{2} &\approx &(4-10)\times 10^{-6}\;{\rm %
eV^{2}}  \label{smallmsw} \\
\sin^{2}2\theta _{\odot } &\approx &(0.1-1.0)\times 10^{-2}
\end{eqnarray}
(ii) a large mixing angle (LMA): 
\begin{eqnarray}
\Delta m_{\odot }^{2} &\approx &(2-20)\times 10^{-5}\;{\rm %
eV^{2}} \\
\sin^{2}2\theta _{\odot } &\approx &(0.65-0.97)
\end{eqnarray}
(iii) vacuum oscillations (VO): 
\begin{eqnarray}
\Delta m_{\odot }^{2} &\approx &(0.5-5)\times 10^{-10}\;%
{\rm eV^{2}} \\
\sin^{2}2\theta_{\odot } &\geq &0.67
\end{eqnarray} 

Clearly these data do not match what would have been the most simple
philosophy: namely, that there be only the SM scales and the 
Planck scale. Putting see-saw particles \cite{seesaw} only with Planck 
masses and assuming all couplings of order unity could not give 
neutrino masses corresponding to the observed oscillations. Rather there 
is a call for a new scale; either some new Higgs field giving neutrino masses 
directly, or a new scale for see-saw particles at $10^{12}~\GeV$.

We are basically forced to take this new scale, say, the see-saw scale, 
as a parameter which is just fitted to neutrino masses and their mixing 
angles. Thus we can, in fact, only hope to make predictions for the 
ratio of the two mass differences observed:

\begin{equation}
  \label{eq:data}
\frac{\Delta m^2_{\odot}}{\Delta m^2_{\rm atm}} = \left\{ \begin{array}{r@{\hspace{1cm}}l}
       0.67-5.0  \times 10^{-3}   & \mbox{for small mixing angle} \\
       0.33-10 \times 10^{-2}   & \mbox{for large mixing angle} \\
       0.08-2.5\times 10^{-7}   & \mbox{for vacuum oscillation} %
\end{array} \right. \end{equation}

\subsection{Content of present article}

\indent

Our goal is to understand the orders of magnitude - only 
orders of magnitude - of masses and mixing angles 
using a model with all small numbers coming from Higgs vacuum 
expectation values (VEV) which are small relative to, say, 
the Planck scale. In the whole paper we do not assume supersymmetry (SUSY), 
but much of our work would not be disturbed by it. However, our work 
is really an extension of an earlier model which we call the ``old'' 
Anti-GUT model - put forward by C.D.~Froggatt and one of us, mainly fitting 
the quark and charged lepton mass matrices - and in these fits the 
same field and its Hermitian conjugate field are both used, and it goes 
into the predictive power that they have the same VEV numerically. In 
SUSY this causes problems of the sort one would have in 
introducing a $\tan\beta$, which would be a new parameter and thus take 
away some predictive power out of the fits, unless  
$\tan\beta\approx 1$. 

Since, however, the neutrino masses only 
come from the up-type Weinberg-Salam Higgs field in SUSY 
models the neutrino mass relations would not be 
disturbed by SUSY. 

The fitting mentioned as the ``old'' Anti-GUT of the charged 
quark and lepton masses and quark mixing angles already presents  
a rather good fit, order-of-magnitude-wise.

However, extending it has met with some difficulty: 
at first, a philosophy of having in addition to the weak 
and the strong scale only the Planck scale would never give 
neutrinos masses in the observed range. So, either lower 
than Planck mass scale see-saw particles are needed, or 
a new, very low VEV Higgs field must be included in the 
model. The latter is the approach of 
M.~Gibson~\etal~\cite{markthesis,mark1,mark2}, who 
introduce a Higgs field triplet under the SM $SU(2)$ with 
weak hypercharge $y/2=1$ to deliver masses for the neutrinos.

The best working model of this type achieves the large
atmospheric neutrino oscillation mixing angle $\theta_{\rm atm}$
by letting it come dominantly from the transformation diagonalising 
the charged lepton mass matrix. A viable scheme is obtained by 
introducing a couple of extra Higgs fields. 

It is the purpose of the present article to seek to extend 
and rescue the ``old'' Anti-GUT model from being falsified
by neutrino oscillations along the first-mentioned road:
introduction of the see-saw particles much lighter than 
the Planck mass; to be more precise, in the range around 
$10^{12}~\GeV$. 

Since this needed scale for the see-saw particles is a
rather ``new'' scale - in the middle of the desert we shall
in the present article - where all small numbers should be 
explained by Higgs field VEVs - just ``shamelessly'' introduce
a new Higgs field $\phi_{\sss B-L}$ which has a VEV of this
size, just fitting it.

Our point is therefore not to explain the overall size 
of neutrino masses, but rather ratio(s) and mixing angles. So 
the ``holy grail'' for this article becomes the ratio of 
the mass square difference describing respectively the
solar neutrino oscillations and the $\nu_{\mu}-\nu_{\tau}$
atmospheric oscillations (\ref{eq:data}).

In the following section, we review briefly the 
Anti-GUT model and its extension to the calculation of  
neutrino mass square differences and mixing angles. Then, 
in the next section, we shall write down the 
no-anomaly constraints on the possible charge assignments for
the quarks and leptons, and present the charge assignments 
for both fermions and Higgs fields in the extended Anti-GUT model.
In section \ref{sec:massmatrizen} we put forward the mass 
matrices and in section \ref{sec:mixingangle} we define the 
mixing angles. Then in section \ref{sec:berechnung} we 
calculate in the model the neutrino oscillation parameters, 
both crudely and numerically, using 
``random order unity factors''. Section \ref{sec:conclusion}
contains our resum{\'e} and conclusion.

\section{Anti-Grand Unification Theory and its extension}
\label{sec:agut}

In this section we review the Anti-GUT model and its 
extension describing neutrino masses and mixing angles.

The Anti-GUT model \cite{mark1,mark2,fn,glasgow,fn2,smg3m,fn3}
has been put forward by one of us and his collaborators
over many years, with several motivations, first of all 
justified by a very promising series of experimental agreements by fitting 
many of the SM parameters with rather few parameters, in 
an impressive way even though most predictions are only order of 
magnitude wise.
\smallskip
The Anti-GUT model deserves its name in as far as its gauge group 
$SMG^3\times U(1)_f$ which, so to speak, replaces the often-used GUT 
gauge groups such as $SU(5)$, $SO(10)$ {\it etc.} by one that can 
be specified by requiring that: %
\begin{enumerate}
\item It should only contain transformations which change the known 
45 (= 3 generations of 15 Weyl particles each) Weyl fermions - counted 
as left-handed, say, - into each other unitarily, ({\it i.e.} it must 
be a subgroup of $U(45)$).
\item It should be anomaly-free even without using the 
Green-Schwarz \cite{gsa} anomaly cancellation mechanism.
\item It should NOT unify the irreducible representations under the 
SM gauge group, called here $SMG$ = $SU(3) \times SU(2) \times U(1)$.
\item It should be as big as possible under the foregoing assumptions.
\end{enumerate}

\smallskip

In the present article we shall, however, allow for see-saw neutrinos
- essentially right-handed neutrinos - whereby we want to extend the 
number of particles to be transformed under the group being specified 
to also include the right-handed neutrinos, even though they have not been
directly ``seen''.

\smallskip

The group which shall be used as the model gauge group replacing the 
unifying groups could be specified by a similar set of assumptions 
like that used by C.D.~Froggatt and one of us to specify the 
``old'' Anti-GUT, by replacing assumption $1$ by a 
slightly modified assumption excluding only already observed fermions in the 
system to only exclude the unobserved fermions when they have nontrivial 
quantum numbers under the SM group, so that they are mass-protected. 
The particles that are mass protected under the SM would namely be rather 
light and would likely have been seen. But see-saw neutrinos with
zero SM quantum numbers could not be mass protected by 
the SM and could easily be so heavy as not to have been ``seen''.

The model which we have in mind as the extended Anti-GUT model, that should 
inherit the successes of the ``old'' Anti-GUT model and in addition have 
see-saw neutrinos, is proposed to have the gauge group
$SMG^3\times U(1)_f\times U(1)_{\sss\rm B-L,1} \times U(1)_{\sss\rm B-L,23}
\!\ni\!SU(3)^3\times SU(2)^3 \times U(1)^6$, and it is assumed to couple 
in the following way:

\noindent
The three SM groups $SMG = SU(3) \times SU(2) \times U(1)$
are supposed to be one for each generation. That is to say, there is, 
{\it e.g.}, a first generation $SMG$ among the three; all the 
fermions in the second and third generations are in the 
trivial representations, and with zero charge, while the 
first generation particles couple to this first generation $SMG$ 
as if in the same representations (same charges too) as they are under 
the SM. For example, the left proto-electron and the proto-electron 
neutrino form a doublet under the $SU(2)_1$ belonging to the first 
generation (while they are in singlets w.r.t. the other $SU(2)$'s) 
and have weak hypercharge w.r.t. the first generation $U(1)_1$, the 
charge $y_1/2 = -1/2$ analogous to the SM weak hypercharge 
being $y/2=-1/2$ for left-handed leptons.

The $U(1)_f$-charge is assigned in a slightly complicated way which is,
however, largely the only one allowed modulo various permutations and 
rewritings from the no-anomaly requirements. It is zero for all 
first-generation and for all particles usually called left-handed, and 
the charge values are the opposite on a ``right-handed'' particle 
in the second generation and the corresponding one in 
the third generation. See Table $1$ for the detailed assignment.

The two last $U(1)$-groups, $U(1)_{\sss\rm B-L,1}$ and 
$U(1)_{\sss\rm B-L,23}$ in our model have charge assignments 
corresponding to the quantum number $B-L$ (= baryon number 
minus lepton number) though in such a way that the charges 
of $U(1)_{\sss\rm B-L,1}$ are zero for the second and third 
generations, and only non-zero for the first generation, for which 
they then coincide with the baryon number minus the 
lepton number. Analogously the 
$U(1)_{\sss\rm B-L,23}$-charge assignments are zero on the 
first-generation quarks and leptons, while they coincide 
with the baryon number minus the lepton number for second and third 
generations. We will discuss in the next section the anomaly 
cancellation in the extended Anti-GUT model.
 
It is then further part of our model that this large gauge group
is broken down spontaneously to the SM group, lying as the 
diagonal subgroup of the $SMG^3$ part of the group by means of a series 
of Higgs fields, the quantum numbers of which have been 
selected mainly from the criteria of fitting the the masses and mixing 
angles w.r.t. order of magnitude. The quantum numbers of the ``old'' 
Anti-GUT Higgs fields proposed were:
\noindent
\begin{eqnarray}
&S:& \quad (\frac{1}{6},-\frac{1}{6},0,-1)\\
&W:& \quad (0,-\frac{1}{2},\frac{1}{2},-\frac{4}{3})\\
&T:& \quad (0,-\frac{1}{6},\frac{1}{6}, -\frac{2}{3})\\
&\xi:& \quad (\frac{1}{6},-\frac{1}{6}, 0,0)
\end{eqnarray}%
\noindent
for the four Higgs fields supposed to have VEVs of 
the order of between a twentieth and unity compared to the fundamental 
scale supposed to be the Planck scale. In addition there was then the 
Higgs field under the Anti-GUT-group which should take the role of finally 
breaking the SM gauge group down to 
$U(3)\!=\! SU(3)\times U(1)_{em}$, {\it i.e.}, play the role 
of the Weinberg-Salam Higgs field
\begin{equation}
\phi_{WS}: \quad (0, \frac{2}{3},-\frac{1}{6},1).
\end{equation} %
\indent
Here the quantum numbers were presented in the order of first giving the 
three different weak hypercharges corresponding to the three generations
$SU(2)_i$, $SU(3)_i$ and $y_i/2 \,(i=1,2,3)$, and then the $U(1)_f$-charge.

In reference \cite{mark1} we fitted the parameters being Higgs fields VEVs
to masses and mixing angles for charged fermions and the values 
are as follows:
\begin{equation}
  \label{eq:vevs}
  \sVEV{S} = 1 \nn,\nn 
  \sVEV{W} = 0.179 \nn,\nn
  \sVEV{\xi} = 0.099 \nn,\nn
  \sVEV{T} = 0.071
\end{equation}

In the following we shall often abbreviate by deleting 
the $\sVEV{\cdots}$ around these Higgs fields, 
mostly with the understanding that $S$, $W$, $\dots$ 
then mean the VEV ``measured in fundamental'' units.

In the Anti-GUT model, old as well as new, it is assumed 
at some ``fundamental scale'' that particles which can play 
the role of see-saw with whatever quantum numbers are needed, 
exist. The fitted ``suppression factors'' are 
the VEVs in units of the ``fundamental scale'' see-saw particles.

It has to be checked that extending the group to have the 
$U(1)_{\sss\rm B-L,1}$ and  $U(1)_{\sss\rm B-L,23}$ does not 
disturb the model already functioning rather well, and it can be done by 
only giving the field $\xi$ and $S$ non-zero charges under 
these ``new'' $U(1)$ groups, so as to get:
\begin{eqnarray}
&S:&  \quad (\frac{1}{6},-\frac{1}{6},0,-1,-\frac{2}{3},\frac{2}{3})\\
&\xi:& \quad (\frac{1}{6},-\frac{1}{6},0,0,\frac{1}{3},-\frac{1}{3})
\end{eqnarray}%

But now we also want to introduce two new Higgs fields $\phi_{\sss B-L}$
and $\chi$ into the model: the first, $\phi_{\sss B-L}$, is a Higgs field 
to fit the new scale that comes in by neutrino oscillations giving the 
scale of the see-saw particle masses. When the left-right-transition 
mass matrix is of the same order as the usual charge fermion mass 
matrices, this scale is of the order $10^{12}~\GeV$. 

We use in our model the gauged $B-L$, in fact the total one, 
because we break $U(1)_{\sss B-L,1}\times U(1)_{\sss B-L,23}%
\supseteq U(1)_{{\sss B-L,}{\rm total}}$ at a much higher 
scale (near Planck scale), to mass-protect the 
right-handed neutrinos meant to function as see-saw particles, so 
they can be sufficiently light to give by the see-saw mechanism 
the ``observed'' left-handed neutrino masses.  The breaking of the 
$U(1)_{{\sss B-L,}{\rm total}}$ and thereby the setting of the 
see-saw scale is then caused by our ``new'' Higgs
field called $\phi_{\sss B-L}$.

In order to get viable neutrino spectra we shall 
choose the quantum numbers of $\phi_{\sss B-L}$ so that 
it is the effective 
$\Bar{\nu_{\tau_{\sss R}}}\,C\,\Bar{\nu_{e_{\sss R}}}^t + h.c.$ 
which gets the direct contribution and thus is not further suppressed.

This is the way to avoid ``factorised mass matrices'' - {\it i.e.} of the 
form
\begin{equation}
  \label{eq:factrization}
\left ( \begin{array}{ccc}
        \phi_{1}^2 &\phi_{1}\phi_{2} & \phi_{1}\phi_{3}\\
 \phi_{1}\phi_{2} &\phi_{2}^2 & \phi_{2}\phi_{3}\\
 \phi_{1}\phi_{3} &\phi_{2}\phi_{3} & \phi_{3}^2
                        \end{array} \right ) 
\end{equation}%
with different order unity factors, though on different 
terms. Such factorised matrices are rather difficult to 
avoid otherwise. If we get such a ``factorised matrix'' and, as 
in our model, have mainly diagonal elements in the $\nu$-Dirac 
matrix, $M_{\nu}^D$, the prediction comes out that 
\begin{equation}
\frac{\Delta m^2_{\odot}}{\Delta m^2_{\rm atm}} \approx (\sin\theta_{\rm atm})^4\nn,  
\end{equation}%
which is not true experimentally. Therefore we choose $\phi_{\sss B-L}$ 
to have the quantum numbers of $\Bar{\nu_{\tau_{\sss R}}}$ plus those of 
$\Bar{\nu_{e_{\sss R}}}$:
\begin{eqnarray}
  \label{eq:blladung}
  Q_{\phi_{\sss B-L}} &=& Q_{\bar{\nu}_{\tau_{\sss R}}} + Q_{\bar{\nu}_{e_{\sss R}}} \nonumber\\
                      &=& (0,0,0,0,1,0) + (0,0,0,1,0,1) \nonumber\\
                      &=& (0,0,0,1,1,1)\nn. \nonumber
\end{eqnarray}
\indent
The other ``new'' relative to ``old'' Anti-GUT Higgs field we call 
$\chi$ and one of its roles is to help
the $\sVEV{\phi_{\sss B-L}}$ to give non-zero effective mass terms for the 
see-saw neutrinos by providing a transition between $\nu_{\tau_{\sss R}}$ 
and $\nu_{\mu_{\sss R}}$. It also turns out to play a role in fitting 
the atmospheric mixing angle (to be of order unity). Its quantum numbers 
are therefore postulated to be the difference of those of these two 
see-saw particles
\begin{eqnarray}
  \label{eq:chiladung}
  Q_{\chi} &=& Q_{\bar{\nu}_{\mu_{\sss R}}} - Q_{\bar{\nu}_{\tau_{\sss R}}} \nonumber\\
           &=& (0,0,0,1,0,-1) - (0,0,0,-1,0,-1) \nonumber\\
           &=& (0,0,0,2,0,0)\nn. \nonumber
\end{eqnarray}
\section{Anomaly}
\label{sec:anomaly}
\indent
We should introduce here an anomaly-free Abelian 
extension of the SM which we shall discuss below
to obtain the neutrino mass spectra and their mixing angles.
The ``new'' Anti-GUT gauge group is
\begin{equation}
  \label{eq:agutgg}
  SMG^3\times U(1)_f\times U(1)_{\sss B-L,1}\times U(1)_{\sss B-L,23} 
\end{equation} and is broken by a set of Higgs fields 
$S$, $W$, $T$, $\xi$, $\chi$ and $\phi_{\sss B-L}$ down to the 
SM gauge groups. The $SMG$ will be broken down by the field
$\phi_{\sss WS}$ playing the role of Weinberg-Salam 
Higgs field into $SU(3)\times U(1)_{em}$.

The requirement that all anomalies involving $U(1)_f$, 
$U(1)_{\sss\rm B-L,1}$ and $U(1)_{\sss\rm B-L,23}$ 
then vanish strongly constrains the possible fermion charges (denoting 
the $U(1)_f$ charges by $Q_f(t_{\sss R}) \equiv t_R$ {\it etc.} and 
the $U(1)_{\sss\rm B-L}$
charges by $Q_{\sss\rm B-L,1}(u_{\sss R}) \equiv \bar{u}_{\sss R}$, 
$Q_{\sss\rm B-L,23}(t_{\sss R}) \equiv \tilde{t}_{\sss R}$ 
{\it etc.} respectively). 

The anomaly cancellation conditions then constrain the fermion 
$U(1)_f$ and $U(1)_{\sss B-L,1}$ and also  $U(1)_{\sss\rm B-L,23}$
charges to satisfy the following equations:
\begin{eqnarray}
\Tr\;[SU_1(3)^2 U(1)_f] &=& 2u_{\sss L}-u_{\sss R}-d_{\sss R} =0 \nonumber \\
\Tr\;[SU_2(3)^2 U(1)_f] &=& 2c_{\sss L}-c_{\sss R}-s_{\sss R} =0 \nonumber \\
\Tr\;[SU_3(3)^2 U(1)_f] &=& 2t_{\sss L}-t_{\sss R}-b_{\sss R} =0 \nonumber \\
\Tr\;[SU_1(2)^2 U(1)_f] &=& 3u_{\sss L}+e_{\sss L} =0 \nonumber \\
\Tr\;[SU_2(2)^2 U(1)_f] &=& 3c_{\sss L}+\mu_{\sss L} =0 \nonumber \\
\Tr\;[SU_3(2)^2 U(1)_f] &=& 3t_{\sss L}+\tau_{\sss L} =0 \nonumber \\
\Tr\;[U_1(1)^2 U(1)_f] &=& u_{\sss L} -8u_{\sss R}-2d_{\sss R}+3e_{\sss L}
                         - 6e_{\sss R} =0 \nonumber \\
\Tr\;[U_2(1)^2 U(1)_f] &=& c_{\sss L}-8c_{\sss R}-2s_{\sss R}+3\mu_{\sss L}
                         -6\mu_{\sss R} =0\nonumber \\
\Tr\;[U_3(1)^2 U(1)_f] &=& t_{\sss L}-8t_{\sss R}-2b_{\sss R}+3\tau_{\sss L}
                         -6\tau_{\sss R} =0\nonumber \\
\Tr\;[U_1(1) U(1)_f^2] &=& u_{\sss L}^2-2u_{\sss R}^2+d_{\sss R}^2-e_{\sss L}^2
                         +e_{\sss R}^2 =0\nonumber \\
\Tr\;[U_2(1) U(1)_f^2] &=& c_{\sss L}^2-2c_{\sss R}^2+s_{\sss R}^2-\mu_{\sss L}^2
                         +\mu_{\sss R}^2 =0\nonumber \\
\Tr\;[U_3(1) U(1)_f^2] &=& t_{\sss L}^2-2t_{\sss R}^2+b_{\sss R}^2-\tau_{\sss L}^2
                         +\tau_{\sss R}^2=0 \nonumber \\
\Tr\;[U(1)_f^3] &=& 6u_{\sss L}^3+6c_{\sss L}^3+6t_{\sss L}^3-3u_{\sss R}^3-3c_{\sss R}^3
                  -3t_{\sss R}^3-3d_{\sss R}^3-3s_{\sss R}^3 \nonumber \\&& 
                  -3b_{\sss R}^3+2e_{\sss L}^3+2\mu_{\sss L}^3 +2\tau_{\sss L}^3
                   -e_{\sss R}^3-\mu_{\sss R}^3-\tau_{\sss R}^3 \nonumber\\ 
           && -\nu_{e_{\sss R}}^3-\nu_{\mu_{\sss R}}^3-\nu_{\tau_{\sss R}}^3 = 0 \nonumber \\
\Tr\;[{(\rm graviton)}^2 U(1)_f] &=& 6u_{\sss L}+6c_{\sss L}+6t_{\sss L}-3u_{\sss R}
-3c_{\sss R}-3t_{\sss R}-3d_{\sss R}-3s_{\sss R} \nonumber \\
&& -3b_{\sss R}+2e_{\sss L}+2\mu_{\sss L}+2\tau_{\sss L}-e_{\sss R}-\mu_{\sss R}
-\tau_{\sss R}\nonumber\\
&& -\nu_{e_{\sss R}}-\nu_{\mu_{\sss R}}-\nu_{\tau_{\sss R}}=0 \nonumber
\end{eqnarray}%
So they should be obeyed both by the $U(1)_{\sss B-L,1}$, and 
$U(1)_{\sss B-L,23}$, replacing the $t_{\sss R}$, $b_{\sss R}$, $\dots$ by 
$\tilde{t}_{\sss R}$, $\tilde{b}_{\sss R}$, $\dots${}:
\begin{eqnarray}
\Tr\;[SU_1(3)^2 U(1)_{\rm\sss B-L,1}] &=& 2\bar{u}_{\sss L}-\bar{u}_{\sss R}-\bar{d}_{\sss R} =0 \nonumber \\
\Tr\;[SU_2(3)^2 U(1)_{\rm\sss B-L,23}] &=& 2\tilde{c}_{\sss L}-\tilde{c}_{\sss R}-\tilde{s}_{\sss R} =0 \nonumber \\
\Tr\;[SU_3(3)^2 U(1)_{\rm\sss B-L,23}] &=& 2\tilde{t}_{\sss L}-\tilde{t}_{\sss R}-\tilde{b}_{\sss R} =0 \nonumber \\
\Tr\;[SU_1(2)^2 U(1)_{\rm\sss B-L,1}] &=& 3\bar{u}_{\sss L}+\bar{e}_{\sss L} =0 \nonumber \\
\Tr\;[SU_2(2)^2 U(1)_{\rm\sss B-L,23}] &=& 3\tilde{c}_{\sss L}+\tilde{\mu}_{\sss L} =0 \nonumber \\
\Tr\;[SU_3(2)^2 U(1)_{\rm\sss B-L,23}] &=& 3\tilde{t}_{\sss L}+\tilde{\tau}_{\sss L} =0 \nonumber
\end{eqnarray}
But with several $U(1)$s there will in addition be anomaly conditions 
for combinations between the different ones. Taking it that 
$U(1)_{\sss B-L,1}$ charges are zero for all second- and third-generation 
fermions while $U(1)_{\sss B-L,23}$ charges are zero for the first generation, the further conditions are:
\begin{eqnarray}
\Tr\;[U_1(1)^2 U(1)_{\rm\sss B-L,1}] &=& \bar{u}_{\sss L} -8\bar{u}_{\sss R}-2\bar{d}_{\sss R}+3\bar{e}_{\sss L}
                         - 6\bar{e}_{\sss R} =0 \nonumber \\
\Tr\;[U_2(1)^2 U(1)_{\rm\sss B-L,23}] &=& \tilde{c}_{\sss L}-8\tilde{c}_{\sss R}-2\tilde{s}_{\sss R}+3\tilde{\mu}_{\sss L} -6\tilde{\mu}_{\sss R} =0\nonumber \\
\Tr\;[U_3(1)^2 U(1)_{\rm\sss B-L,23}] &=& \tilde{t}_{\sss L}-8\tilde{t}_{\sss R}-2\tilde{b}_{\sss R}
+3\tilde{\tau}_{\sss L}-6\tilde{\tau}_{\sss R} =0\nonumber \\
\Tr\;[U_1(1) U(1)_{\rm\sss B-L,1}^2] &=& \bar{u}_{\sss L}^2-2\bar{u}_{\sss R}^2+\bar{d}_{\sss R}^2-\bar{e}_{\sss L}^2
                         +\bar{e}_{\sss R}^2 =0\nonumber \\
\Tr\;[U_2(1) U(1)_{\rm\sss B-L,23}^2] &=& \tilde{c}_{\sss L}^2-2\tilde{c}_{\sss R}^2+\tilde{s}_{\sss R}^2-\tilde{\mu}_{\sss L}^2+\tilde{\mu}_{\sss R}^2 =0\nonumber \\
\Tr\;[U_3(1) U(1)_{\rm\sss B-L,23}^2] &=& \tilde{t}_{\sss L}^2-2\tilde{t}_{\sss R}^2+\tilde{b}_{\sss R}^2-\tilde{\tau}_{\sss L}^2+\tilde{\tau}_{\sss R}^2=0 \nonumber \\
\Tr\;[U(1)_{\rm\sss B-L,1}^3] &=& 6\bar{u}_{\sss L}^3-3\bar{u}_{\sss R}^3-3\bar{d}_{\sss R}^3+2\bar{e}_{\sss L}^3-\bar{e}_{\sss R}^3-\bar{\nu}_{e_{\sss R}}^3=0 \nonumber \\
\Tr\;[U(1)_{\rm\sss B-L,23}^3] &=& 6\tilde{c}_{\sss L}^3+6\tilde{t}_{\sss L}^3-3\tilde{c}_{\sss R}^3-3\tilde{t}_{\sss R}^3-3\tilde{s}_{\sss R}^3-3\tilde{b}_{\sss R}^3\nonumber\\
&&  +2\tilde{\mu}_{\sss L}^3 +2\tilde{\tau}_{\sss L}^3-\tilde{\mu}_{\sss R}^3-\tilde{\tau}_{\sss R}^3-\tilde{\nu}_{\mu_{\sss R}}^3-\tilde{\nu}_{\tau_{\sss R}}^3=0 \nonumber \\
\Tr\;[{(\rm graviton)}^2 U(1)_{\rm\sss B-L,1}] &=& 6\bar{u}_{\sss L}-3\bar{u}_{\sss R}-3\bar{d}_{\sss R}+2\bar{e}_{\sss L}-\bar{e}_{\sss R}-\bar{\nu}_{e_{\sss R}}=0 \nonumber\\
\Tr\;[{(\rm graviton)}^2 U(1)_{\rm\sss B-L,23}] &=& 6\tilde{c}_{\sss L}+6\tilde{t}_{\sss L}-3\tilde{c}_{\sss R}-3\tilde{t}_{\sss R}-3\tilde{s}_{\sss R}-3\tilde{b}_{\sss R}\nonumber\\
&&  +2\tilde{\mu}_{\sss L}+2\tilde{\tau}_{\sss L}-\tilde{\mu}_{\sss R}-\tilde{\tau}_{\sss R}-\tilde{\nu}_{\mu_{\sss R}}-\tilde{\nu}_{\tau_{\sss R}}=0 \nonumber\\
\Tr\;[U(1)_f^2 U(1)_{\sss B-L,1}] &=&  
6u_{\sss L}^2\bar{u}_{\sss L}-3u_{\sss R}^2\bar{u}_{\sss R}-3d_{\sss R}^2\bar{d}_{\sss R} %
+2e_{\sss L}^2\bar{e}_{\sss L}- e_{\sss R}^2\bar{e}_{\sss R} -\nu_{e_{\sss R}}^2\bar{\nu}_{e_{\sss R}} = 0\nonumber\\
\Tr\;[U(1)_f^2 U(1)_{\sss B-L,23}] &=&  6c_{\sss L}^2\tilde{c}_{\sss L}-3c_{\sss R}^2\tilde{c}_{\sss R}-3s_{\sss R}^2\tilde{s}_{\sss R}%
+ 2\mu_{\sss L}^2\tilde{\mu}_{\sss L}- \mu_{\sss R}^2\tilde{\mu}_{\sss R} -\nu_{\mu_{\sss R}}^2\tilde{\nu}_{\mu_{\sss R}}\nonumber\\
&& +6t_{\sss L}^2\tilde{t}_{\sss L}-3t_{\sss R}^2\tilde{t}_{\sss R} -3b_{\sss R}^2\tilde{b}_{\sss R} %
+ 2\tau_{\sss L}^2\tilde{\tau}_{\sss L} - \tau_{\sss R}^2\tilde{\tau}_{\sss R} 
-\nu_{\tau_{\sss R}}^2\tilde{\nu}_{\tau_{\sss R}} = 0\nonumber\\
\Tr\;[U(1)_f U(1)_{\sss B-L,1}^2] &=&  
6u_{\sss L}\bar{u}_{\sss L}^2-3u_{\sss R}\bar{u}_{\sss R}^2-3d_{\sss R}\bar{d}_{\sss R}^2 %
+2e_{\sss L}\bar{e}_{\sss L}^2- e_{\sss R}\bar{e}_{\sss R}^2 -\nu_{e_{\sss R}}\bar{\nu}_{e_{\sss R}}^2 = 0\nonumber\\
\Tr\;[U(1)_f U(1)_{\sss B-L,23}^2] &=&  6c_{\sss L}\tilde{c}_{\sss L}^2-3c_{\sss R}\tilde{c}_{\sss R}^2-3s_{\sss R}\tilde{s}_{\sss R}^2%
+ 2\mu_{\sss L}\tilde{\mu}_{\sss L}^2- \mu_{\sss R}\tilde{\mu}_{\sss R}^2 -\nu_{\mu_{\sss R}}\tilde{\nu}_{\mu_{\sss R}}^2\nonumber\\
&& +6t_{\sss L}\tilde{t}_{\sss L}^2-3t_{\sss R}\tilde{t}_{\sss R}^2 -3b_{\sss R}\tilde{b}_{\sss R}^2 %
+ 2\tau_{\sss L}\tilde{\tau}_{\sss L}^2 - \tau_{\sss R}\tilde{\tau}_{\sss R}^2 
-\nu_{\tau_{\sss R}}\tilde{\nu}_{\tau_{\sss R}}^2 = 0\nonumber\\
\Tr\;[U(1)_1 U(1)_f U(1)_{\sss B-L,1}] &=& u_{\sss L}\bar{u}_{\sss L} 
-2 u_{\sss R}\bar{u}_{\sss R} + d_{\sss R}\bar{d}_{\sss R} - e_{\sss L}\bar{e}_{\sss L}+ e_{\sss R}\bar{e}_{\sss R} = 0\nonumber\\
\Tr\;[U(1)_2 U(1)_f U(1)_{\sss B-L,23}] &=&  c_{\sss L}\tilde{c}_{\sss L} 
-2 c_{\sss R}\tilde{c}_{\sss R} + s_{\sss R}\tilde{s}_{\sss R}- \mu_{\sss L}\tilde{\mu}_{\sss L}+ \mu_{\sss R}\tilde{\mu}_{\sss R} = 0\nonumber\\
\Tr\;[U(1)_3 U(1)_f U(1)_{\sss B-L,23}] &=&  t_{\sss L}\tilde{t}_{\sss L} 
-2 t_{\sss R}\tilde{t}_{\sss R} + b_{\sss R}\tilde{b}_{\sss R} - \tau_{\sss L}\tilde{\tau}_{\sss L}+ \tau_{\sss R}\tilde{\tau}_{\sss R} = 0\nonumber
\end{eqnarray}

\vspace*{.5cm}

\begin{table}[!tt]
\caption{All $U(1)$ quantum charges in extended Anti-GUT model}
\label{Table1}
\begin{center}
\begin{tabular}{|c||c|c|c|c|c|c|} \hline
& $SMG_1$& $SMG_2$ & $SMG_3$ & $U(1)_f$ & $U_{\sss B-L,1}$ & $U_{\sss B-L,23}$ \\ \hline\hline
$u_L,d_L$ &  $\frac{1}{6}$ & $0$ & $0$ & $0$ & $\frac{1}{3}$ & $0$ \\
$u_R$ &  $\frac{2}{3}$ & $0$ & $0$ & $0$ & $\frac{1}{3}$ & $0$ \\
$d_R$ & $-\frac{1}{3}$ & $0$ & $0$ & $0$ & $\frac{1}{3}$ & $0$ \\
$e_L, \nu_{e_{\sss L}}$ & $-\frac{1}{2}$ & $0$ & $0$ & $0$ & $-1$ & $0$ \\
$e_R$ & $-1$ & $0$ & $0$ & $0$ & $-1$ & $0$ \\
$\nu_{e_{\sss R}}$ &  $0$ & $0$ & $0$ & $0$ & $-1$ & $0$ \\ \hline
$c_L,s_L$ & $0$ & $\frac{1}{6}$ & $0$ & $0$ & $0$ & $\frac{1}{3}$ \\
$c_R$ &  $0$ & $\frac{2}{3}$ & $0$ & $1$ & $0$ & $\frac{1}{3}$ \\
$s_R$ & $0$ & $-\frac{1}{3}$ & $0$ & $-1$ & $0$ & $\frac{1}{3}$\\
$\mu_L, \nu_{\mu_{\sss L}}$ & $0$ & $-\frac{1}{2}$ & $0$ & $0$ & $0$ & $-1$\\
$\mu_R$ & $0$ & $-1$ & $0$ & $-1$  & $0$ & $-1$ \\
$\nu_{\mu_{\sss R}}$ &  $0$ & $0$ & $0$ & $1$ & $0$ & $-1$ \\ \hline
$t_L,b_L$ & $0$ & $0$ & $\frac{1}{6}$ & $0$ & $0$ & $\frac{1}{3}$ \\
$t_R$ &  $0$ & $0$ & $\frac{2}{3}$ & $-1$ & $0$ & $\frac{1}{3}$ \\
$b_R$ & $0$ & $0$ & $-\frac{1}{3}$ & $1$ & $0$ & $\frac{1}{3}$\\
$\tau_L, \nu_{\tau_{\sss L}}$ & $0$ & $0$ & $-\frac{1}{2}$ & $0$ & $0$ & $-1$\\
$\tau_R$ & $0$ & $0$ & $-1$ & $1$ & $0$ & $-1$\\
$\nu_{\tau_{\sss R}}$ &  $0$ & $0$ & $0$ & $-1$ & $0$ & $-1$ \\ \hline \hline
$\phi_{\sss WS}$ & $0$ & $\frac{2}{3}$ & $-\frac{1}{6}$ & $1$ & $0$ & $0$ \\
$S$ & $\frac{1}{6}$ & $-\frac{1}{6}$ & $0$ & $-1$ & $-\frac{2}{3}$ & $\frac{2}{3}$\\
$W$ & $0$ & $-\frac{1}{2}$ & $\frac{1}{2}$ & $-\frac{4}{3}$ & $0$ & $0$ \\
$\xi$ & $\frac{1}{6}$ & $-\frac{1}{6}$ & $0$ & $0$ & $\frac{1}{3}$ & $-\frac{1}{3}$\\
$T$ & $0$ & $-\frac{1}{6}$ & $\frac{1}{6}$ & $-\frac{2}{3}$ & $0$ & $0$\\
$\chi$ & $0$ & $0$ & $0$ & 2 & $0$ & $0$ \\
$\phi_{\sss B-L}$ & $0$ & $0$ & $0$ & $1$ & $1$ & $1$ \\ \hline
\end{tabular}
\end{center}
\end{table}

From these equations we can get the following solutions:
\begin{eqnarray}
(u_{\sss L},u_{\sss R},d_{\sss R},e_{\sss L},e_{\sss R},\nu_{e_{\sss R}}) &=& (0,0,0,0,0,0) \nonumber \\
(c_{\sss L},c_{\sss R},s_{\sss R},\mu_{\sss L},\mu_{\sss R},\nu_{\mu_{\sss R}}) &=& (0,1,-1,0,-1,1) \nonumber \\
(t_{\sss L},t_{\sss R},b_{\sss R},\tau_{\sss L},\tau_{\sss R},\nu_{\tau_{\sss R}}) &=& (0,-1,1,0,1,-1)\nonumber\\
(\bar{u}_{\sss L}, \bar{u}_{\sss R}, \bar{d}_{\sss L}, \bar{d}_{\sss R},\bar{e}_{\sss L}, \bar{e}_{\sss R}, \bar{\nu}_{e_{\sss L}}, \bar{\nu}_{e_{\sss R}}) &=& (\frac{1}{3},\frac{1}{3},\frac{1}{3},\frac{1}{3}, -1,-1,-1,-1) \nonumber\\
(\tilde{c}_{\sss L}, \tilde{c}_{\sss R}, \tilde{s}_{\sss L}, \tilde{s}_{\sss R},\tilde{b}_{\sss L}, \tilde{b}_{\sss R}, \tilde{t}_{\sss L}, \tilde{t}_{\sss R}) &=& (\frac{1}{3},\frac{1}{3},\frac{1}{3},\frac{1}{3},\frac{1}{3},\frac{1}{3},\frac{1}{3},\frac{1}{3}) \nonumber\\
(\tilde{\mu}_{\sss L}, \tilde{\mu}_{\sss R}, \tilde{\tau}_{\sss L}, \tilde{\tau}_{\sss R},\tilde{\nu}_{\mu_{\sss L}}, \tilde{\nu}_{\mu_{\sss R}}, \tilde{\nu}_{\tau_{\sss L}}, \tilde{\nu}_{\tau_{\sss R}}) &=& (-1,-1,-1,-1,-1,-1,-1,-1)\nonumber
\end{eqnarray}

We summarise the Abelian gauge quantum numbers of our model for 
fermions and scalars in Table $1$. However, the following
three points should be kept in mind; then the information 
in Table $1$ and these three points describe our whole model:
\begin{enumerate}
\item We have only presented here the six $U(1)$-charges in our 
model. The non-Abelian quantum charge numbers are to be derived 
from the following rule:
\smallskip%
find in the table $y_i/2$ ($i=1,2,3$ is the generation number), then 
find that Weyl particle in the SM for which the SM weak hypercharge 
divided by two is $y/2=y_i/2$ and use its $SU(2)$ and $SU(3)$ 
representation for the particle considered in the 
table. But now use it for $SU(2)_i$ and $SU(3)_i$.
\item Remember that we imagine that at the ``fundamental'' scale ($\simeq$ 
presumed to be the Planck scale) we have essentially all particles that can 
be imagined with couplings of order unity. But we do not want to be specific 
about these very heavy particles in order not to decrease the 
enormous likelihood of our model being right. We are only specific 
about the particles in the table and the gauge fields.
\item The $39$ gauge bosons correspond to the group (equation
(\ref{eq:agutgg})) and are also not written in the table.
\end{enumerate}

\section{Mass matrices within the Anti-GUT model}
\label{sec:massmatrizen}

In the ``old'' Anti-GUT model we have only the usual SM fermions at 
low energies, but in our ``new'' version we assume that 
there exist very heavy right-handed neutrinos, 
all of them having already decayed and not being observable in 
our world. They shall function as see-saw particles
and thus give rise to an effective Majorana-type mass matrix for the 
left-handed particles. These three ``right-handed''neutrinos 
get masses from the VEV of $\phi_{\sss B-L}$ $(10^{12}~\GeV)$, $\xi$ and 
also $\chi$ Higgs fields (the latter in order of Planck unit,
about $1/10$).

The effective mass matrix elements, left-left, for the left-handed
neutrinos - the ones we ``see'' experimentally - then come about using
the $\nu_R$ see-saw propagator surrounded by left-right transition 
neutrino mass matrices. The latter are rather analogous to 
the charged lepton and quark mass matrices, which are proportional
to the VEV of the Weinberg-Salam Higgs field in our model, 
being components of $\phi_{\sss WS}$ (with VEV $\sim 173~\GeV$).

\smallskip

Both $B-L$ quantum gauge groups are violated 
by $\phi_{\sss\rm B-L}$, thus the effective Majorana mass terms 
are added into the Lagrange density using the Higgs field 
$\phi_{\sss B-L}$. The part of the effective Lagrangian we 
have to consider is:
\begin{eqnarray}
  \label{lagrangian}
-\cL_{\sss\rm lepton-mass} &\!=\!& 
\bar{\nu}_L \,M^D_\nu \,\nu_R 
+  \frac{1}{2}(\Bar{\nu_L})^{\sss c}\,M_{L}\, \nu_L 
+  \frac{1}{2}(\Bar{\nu_R})^{\sss c}\,M_{R}\, \nu_R + h.c. \nonumber\\
&\!=\!&
\frac{1}{2} (\Bar{n_L})^{\sss c} \,M\, n_L + h.c.
\end{eqnarray}
where 
\begin{equation}
n_L \!\equiv\! \left( \begin{array}{c}
    \nu_L \\
    (\nu_L)^{\sss c}
    \end{array} \right) \nn,\nn
M \!\equiv\! \left( \begin{array}{cc}
    M_L & M^D_\nu\\
    M^D_\nu & M_R 
    \end{array} \right) \nn; 
\end{equation}
\noindent
$M^D_\nu$ is the standard $SU(2)\times U(1)$ breaking Dirac mass 
term, and and $M_L$ and $M_R$ are the isosinglet Majorana mass 
terms of left-handed and right-handed neutrinos, respectively. 

\smallskip

Supposing that the left-handed Majorana mass $M_L$ terms are 
comparatively negligible, because of SM gauge symmetry protection,
a naturally small effective Majorana mass for the light 
neutrinos (predominantly $\nu_{\sss L}$) can be generated by mixing 
with the heavy states (predominantly $\nu_{\sss R}$) of mass 
$M_{\nu_{\sss R}}$. The Dirac mass matrix of neutrinos is  
similar to the up-type quark mass matrix \cite{stech} and therefore 
has similar magnitude. For no left-left term, the light eigenvalues 
of the matrix $M$ are
\begin{equation}
  \label{eq:meff}
  M_{\rm eff} \! \approx \! M^D_\nu\,M_R^{-1}\,(M^D_\nu)^t\nn.
\end{equation}

This result is the well-known see-saw mechanism \cite{seesaw}: the 
light neutrino masses are quadratic in the Dirac masses and 
inversely proportional to the large $\nu_R$ Majorana masses. Notice that
if some $\nu_{\sss R}$ are massless or light they would not
be integrated away but simply added to the light neutrinos.

\bigskip

We have already given the quantum charges of the Higgs fields,
$S$, $W$, $T$, $\xi$, $\phi_{\sss WS}$, $\phi_{\sss B-L}$ and
$\chi$ in Table $1$. With this quantum number choice of Higgs
fields the mass matrices are given by
\noindent%
the uct-quarks:%
\begin{equation}
M_U \simeq \frac{\sVEV{\phi_{\sss\rm WS}}}{\sqrt{2}}\hspace{-0.1cm}
\left ( \begin{array}{ccc}
        S^{\dagger}W^{\dagger}T^2(\xi^{\dagger})^2
        & W^{\dagger}T^2\xi & (W^{\dagger})^2T\xi \\
        S^{\dagger}W^{\dagger}T^2(\xi^{\dagger})^3
        & W^{\dagger}T^2 & (W^{\dagger})^2T \\
        S^{\dagger}(\xi^{\dagger})^3 & 1 & W^{\dagger}T^{\dagger}
                        \end{array} \right ) \label{M_U}
\end{equation}\noindent %
the dsb-quarks:
\begin{equation}
M_D \simeq \frac{\sVEV{\phi_{\sss\rm WS}}}{\sqrt{2}}\hspace{-0.1cm}
\left ( \begin{array}{ccc}
        SW(T^{\dagger})^2\xi^2 & W(T^{\dagger})^2\xi & T^3\xi \\
        SW(T^{\dagger})^2\xi & W(T^{\dagger})^2 & T^3 \\
        SW^2(T^{\dagger})^4\xi & W^2(T^{\dagger})^4 & WT
                        \end{array} \right ) \label{M_D}
\end{equation}\noindent %
the charged leptons:
\begin{equation}
M_E \simeq \frac{\sVEV{\phi_{\sss\rm WS}}}{\sqrt{2}}\hspace{-0.1cm}
\left ( \hspace{-0.2 cm}\begin{array}{ccc}
        SW(T^{\dagger})^2\xi^2 & W(T^{\dagger})^2(\xi^{\dagger})^3
        & WT^4(\xi^{\dagger})^3\chi\\
        SW(T^{\dagger})^2\xi^5 & W(T^{\dagger})^2 &
        WT^4\chi\\
        S(W^{\dagger})^2T^4\xi^5 & (W^{\dagger})^2T^4 & WT
                        \end{array} \hspace{-0.2 cm}\right ) \label{M_E}
\end{equation}\noindent%
the Dirac neutrinos:
\begin{equation}
M^D_\nu \simeq \frac{\sVEV{\phi_{\sss\rm WS}}}{\sqrt{2}}\hspace{-0.1cm}
\left ( \hspace{-0.2 cm}\begin{array}{ccc}
        S^{\dagger}W^{\dagger}T^2(\xi^{\dagger})^2 & %
        W^{\dagger}T^2(\xi^{\dagger})^3
        & (W^\dagger)T^2(\xi^\dagger)^3\chi\\
        S^{\dagger}W^{\dagger}T^2\xi & W^{\dagger}T^2 &
        (W^\dagger)T^2\chi\\
        S^{\dagger}W^{\dagger}T^\dagger\xi\chi^\dagger& 
        W^{\dagger}T^\dagger\chi^\dagger 
        & W^{\dagger}T^{\dagger}
                        \end{array} \hspace{-0.2 cm}\right ) \label{Md_N}
\end{equation}\noindent %
and the Majorana neutrinos:
\begin{equation}
M_R \simeq  \sVEV{\phi_{\sss\rm B-L}}\hspace{-0.1cm}
\left (\hspace{-0.2 cm}\begin{array}{ccc}
S^\dagger\chi^\dagger\xi &  \chi^\dagger & 1 \\
 \chi^\dagger & S\chi^\dagger\xi^\dagger & S\xi^\dagger \\
 1 & S\xi^\dagger & S\chi\xi^\dagger
\end{array} \hspace{-0.2 cm}\right ) \label{Mr_N}
\end{equation}

Note that the random complex order of unity and factorial factors 
which are supposed to multiply all the mass matrix elements 
are not represented here. We will discuss these factors in 
section \ref{sec:berechnung}.

The matrices for the quarks $M_{U}$ and $M_{D}$ happen 
not to have been changed at all by the introduction of the 
``new'' Higgs fields $\chi$ (and $\phi_{\sss B-L}$, but 
that has so little VEV compared to the Planck scale that it could 
never compete), and even in the charged lepton mass matrix 
the appearance of $\chi$ occurs on off-diagonal matrix 
elements which are already small and remain so small as to 
have no significance for the charge lepton mass predictions 
as long as $\chi$ is of the order $\sVEV{\chi}\approx0.07$ as we need 
for fitting $\theta_{\rm atm}$.

Therefore all the fits of the ``old'' Anti-GUT model are valid and we
can still use the parameter values obtained by these earlier fits to
$S$, $W$, $T$, $\xi$, presented above in equation (\ref{eq:vevs}). 

\section{Mixing Angles in extended Anti-GUT}
\label{sec:mixingangle}

The neutrino flavour eigenstates $\nu_{\alpha}$ are related to 
the mass eigenstates $\nu_{i}$ in the vacuum by a unitary 
matrix $U$,
\begin{equation}
  \label{eigen}
  \ket{\nu_{\alpha}} = \sum_i U_{\alpha i} \ket{\nu_i}\nn.
\end{equation}

We will investigate in this paper a three-neutrino-generation 
model, so the Maki-Nakagawa-Sakata (MNS)\cite{mns} mixing 
matrix becomes a $3\times3$ matrix and is parametrised by 
\begin{equation}
U
= \left( \begin{array}{ccc}
  c_{13} c_{12}       & c_{13} s_{12}  & s_{13} e^{-i\delta_{13}} \\
- c_{23} s_{12} - s_{13} s_{23} c_{12} e^{i\delta_{13}}
& c_{23} c_{12} - s_{13} s_{23} s_{12} e^{i\delta_{13}}
& c_{13} s_{23} \\
    s_{23} s_{12} - s_{13} c_{23} c_{12} e^{i\delta_{13}}
& - s_{23} c_{12} - s_{13} c_{23} s_{12} e^{i\delta_{13}}
& c_{13} c_{23} \\
\end{array} \right)\left( \begin{array}{ccc}
 e^{i\alpha} & 0 & 0 \\
 0 &  e^{i\beta} & 0\\
 0 &  0          & 1 \\
\end{array} \right) \nn,
\label{eq:MNS}
\end{equation}
where $c_{ij} \equiv \cos\theta_{ij}$ and 
$s_{ij} \equiv \sin\theta_{ij}$ and $\delta_{13}$ is the CP-violating 
phase. Here two CP-violation Majorana phases $\alpha$, $\beta$ are also 
included.

\smallskip

Since in the parametrisation equation (\ref{eq:MNS}) of the mixing matrix 
the CP-violating phase $\delta_{13}$ is associated with 
$\sin\theta_{13}$, it is clear that CP-violation is 
negligible in the lepton sector if the mixing angle 
$\theta_{13}$ is small and for our estimations we should
not consider them, since we will not discuss CP-violation 
in this paper.

Since in our model it turns out that the main mixing of leptons 
comes from the rotation of the (left-handed) neutrino eigenstates 
relative to the protoflavour eigenstates (by ``protoflavour 
eigenstates'' is understood states of definite $U(1)_f, y_i/2, 
(B-L)_{1,{\rm gen}}, (B-L)_{2,{\rm gen}}$ charges) rather than
from the rotations of the charged mass eigenstates relative to the 
(charged) protoflavour eigenstate (leptons), it is preferable 
to choose a parametrisation such that the mixing angle 
$\sin\theta_{13}$ becomes small in this situation. It is the 
matrix element representing the overlap of the heaviest 
left neutrino mass eigenstate with the $\nu_{e_{\sss L}}$,
the state which couples to $W^\pm$ bosons, which is small, of 
order $\xi^3$; in fact we get $\sqrt{10^{-4}}\approx 10^{-2}$.
So we shall take the parametrisation in which this matrix 
element is just $\sin\theta_{13}\,e^{-i\delta_{13}}$ and not 
$\sin\theta_{23}\sin\theta_{12} - \sin\theta_{13}%
\cos\theta_{23}\cos\theta_{12}\,e^{i\delta_{13}}$, as comes 
out in an alternative parametrisation. But for phenomenology 
today, the most crucial mixing angles are only the 
\begin{eqnarray}
  \label{eq:sinbezieung}
  \sin^22\theta_{\odot} &=& \sin^22\theta_{12} \\
  \sin^22\theta_{\rm atm} &=& \sin^22\theta_{23} \nn,
\end{eqnarray}%
defining
\begin{eqnarray}
  \label{eq:mixingmatrix}
  U_E M_E M_E^\dagger U_E^\dagger &=& {\rm diag}(m^2_{\sss e}, m^2_{\sss \mu},m^2_{\sss \tau}) \\
U_{\rm eff} M_{\rm eff} M_{\rm eff}^\dagger U_{\rm eff}^\dagger &=& {\rm diag}(m^2_{\nu_{\sss e}}, m^2_{\nu_{\sss \mu}},m^2_{\nu_{\sss \tau}})\nn.
\end{eqnarray}

The $(1,3)$-component of the $U_{\rm eff}\,U_E^\dagger$ in our model
turns out to be of the order of magnitude of product 
$\sin\theta_{12}\sin\theta_{23}$, while the $(3,1)$-component 
is smaller and it is natural
to choose the parametrisation corresponding to putting
\begin{equation}
  U = U_E\,U_{\rm eff}^\dagger\,=\,(U^\dagger)_{i\alpha}\nn,
\end{equation}%
where $U$ has a very small $\sin\theta_{13}$ achieved when $U$
is parametrised according to equation (\ref{eq:MNS}).

\section{Calculation of $M_{\rm eff}$}
\label{sec:berechnung}

We calculate in this section the effective neutrino mass matrix for
left-handed components. Since, strictly speaking, our model only
predicts orders of magnitude, a crude calculation is in principle 
justified. This calculation is presented in the first 
subsection, and then in the next subsection we 
make ``statistical calculations'' with random order-one 
factors and ``factorial factors''.

\smallskip

\subsection{Crude calculation}

From equation~(\ref{Mr_N}) we see to the first approximation 
that there are one massless and two degenerate right-handed 
neutrinos coming from the VEV of the $B\!-\!L$ breaking 
Higgs field, $\sVEV{\phi_{\sss\rm B-L}}$.

The splitting between the two almost degenerate see-saw 
neutrinos would be $M_{31}\sVEV{S}\sVEV{\chi}\sVEV{\xi}$, 
where $M_{31}\approx\sVEV{\phi_{\sss\rm B-L}}$ is the approximately 
common mass of the two heaviest see-saw neutrinos. The third lightest 
see-saw neutrino is dominantly ``proto second generation'' 
and has the mass $\sVEV{\phi_{\sss\rm B-L}}\sVEV{\chi}\sVEV{\xi}$.

For the left-handed neutrinos to the first approximation we get 
the effective mass matrix as follows:
\begin{equation}
M_{\rm eff}\approx
\frac{W^2T^2\sVEV{\phi_{\sss WS}}^2}{2\sVEV{\phi_{\sss\rm B-L}}}\left( \begin{array}{ccc}
\frac{T^2\xi^5}{\chi} & \frac{T^2\xi^2}{\chi} & T\xi^2 \\
\frac{T^2\xi^2}{\chi} & \frac{T}{\xi} & \frac{T}{\xi}\\
T\xi^2 &   \frac{T}{\xi} & \frac{\chi}{\xi} \\
\end{array} \right) \nn,
\label{eq:hmatrix}
\end{equation}

But we have to emphasise here that this approximation
is {\em not good enough to calculate only the heaviest left-handed neutrino},
because all the mass matrix elements to this approximation 
come from the propagator contribution of the lightest see-saw 
particle, so that they really form a degenerate matrix of 
rank one.  Using this contribution only would lead to two left-handed 
massless neutrinos and one massive. But we can still obtain the 
mixing angles $\theta_{13}$ and $\theta_{23}$ and the heaviest 
mass from
$M_{\rm eff}$:
\begin{eqnarray}
  \label{eq:mixing1223}
  \theta_{13}&=&\theta_{e{},{\rm heavy}}\approx\frac{T}{\chi} \xi^3 \\
  \theta_{23}&=&\theta_{\mu{},{\rm heavy}}\approx\left\{ \begin{array}{r@{\hspace{1cm}}l}
       \frac{T}{\chi} & \mbox{when $\chi\sgeq T$} \\
       1              & \mbox{when $\chi\sleq T$} \\
\end{array} \right. \\
M_{\sss \nu_{\sss L}{\rm heavy}}\!\!\! &\approx&
\left\{ \begin{array}{r@{\hspace{1cm}}l}
\frac{W^2T^2\sVEV{\phi_{\sss WS}}^2}{2\sVEV{\phi_{\sss\rm B-L}}}
\frac{\chi}{\xi} & \mbox{when $\chi\sgeq T$} \\
\frac{W^2T^2\sVEV{\phi_{\sss WS}}^2}{2\sVEV{\phi_{\sss\rm B-L}}}
\frac{T}{\xi} & \mbox{when $\chi\sleq T$} \\
\end{array} \right. \end{eqnarray}
From these equations we can restrict the region of $\chi$ 
comparing with Super-Kamiokande experimental data; $\chi$ must be
almost of the same order as $T$. Thus we know the mixing angle of 
the first and third generations must be of the order of $\xi^3$. 

\bigskip
\indent
However, to get the much lower masses we cannot use the 
contribution from the lightest see-saw propagator, but we have 
to use the propagator terms from the two approximately equally 
heavy see-saw particles. This contribution to the propagator 
matrix is
\begin{equation}
  \label{eq:invmr}
  M^{-1}_{{\sss R}}{}_{\hspace{-0.5mm}\big|}{}_{\tiny\textrm{{\begin{tabular}{l}\hspace{-1.6mm}heavy\\\hspace{-1.6mm}see-saws\end{tabular}}}}%
\!\!\!\!\approx \frac{1}{\sVEV{\phi_{\sss\rm B-L}}}\hspace{-0.1cm}
\left (\hspace{-0.2 cm}\begin{array}{ccc}
\chi\xi &  \xi & 1 \\
 \xi& \chi\xi & \chi \\
 1 & \chi & \chi\xi
\end{array} \hspace{-0.2 cm}\right )
\end{equation}
where the $\chi\xi/\sVEV{\phi_{\sss B-L}}$ comes from the mass difference 
of the almost degenerate see-saw particles.

Surrounding this propagator contribution with the ``Dirac $\nu$''-mass 
matrix we get
\begin{eqnarray}
  \label{eq:appsee}
M_{{\rm eff}}{}_{\hspace{-0.3mm}\big|}{}_{\tiny\textrm{{\begin{tabular}{l}\hspace{-1.6mm}heavy\\\hspace{-1.6mm}see-saws\end{tabular}}}}%
\!\!\!\!&\approx&  M^D_\nu\,M^{-1}_{{\sss R}}{}_{\hspace{-0.5mm}\big|}{}_{\tiny\textrm{{\begin{tabular}{l}\hspace{-1.6mm}heavy\\\hspace{-1.6mm}see-saws\end{tabular}}}}\!\!(M^D_\nu)^t\nonumber\\
 &\approx&  \frac{W^2T^2\sVEV{\phi_{\sss WS}}^2}{2\sVEV{\phi_{\sss\rm B-L}}}\hspace{-0.1cm}
\left (\hspace{-0.2 cm}\begin{array}{ccc}
T^2\xi^6 &  T\xi^3 & T\xi^2 \\
T\xi^3 & T^2\chi\xi & T\chi \\
 T\xi^2 & T\chi & \chi\xi
\end{array} \hspace{-0.2 cm}\right )\nn.
\end{eqnarray}
It is from this contribution that the two lightest left-handed neutrino 
masses and their mixing angle, $\theta_{12}$, should be obtained:
\begin{eqnarray}
  &&M_{\sss \nu_{\sss L}{\rm medium}}\approx \frac{W^2T^2\chi\xi\sVEV{\phi_{\sss WS}}^2}%
{2\sVEV{\phi_{\sss\rm B-L}}} \\
  &&\theta_{12}=\theta_{e{},{\rm medium}}\approx \left\{ \begin{array}{r@{\hspace{1cm}}l}
       \frac{T}{\chi}\xi & \mbox{when $\chi\sgeq T$} \\
       \xi               & \mbox{when $\chi\sleq T$}\nn. \\
\end{array} \right. 
\end{eqnarray}
Note that the lightest mass is quite dominantly the $\nu_{e_{\sss L}}$ neutrino 
and the small mixing angle goes mainly to the medium mass neutrino (%
$\theta_{e{},{\rm medium}}/\theta_{e{},{\rm heavy}}\approx\xi^{-2}\gg 1$). So 
we should identify approximately the solar oscillation mixing angle with 
the mixing to the medium heavy neutrino:
\begin{equation}
  \label{eq:solxi}
  \theta_{\odot}\simeq \theta_{e{},{\rm medium}}\approx \xi 
\end{equation}%
and the solar mass square difference 
\begin{equation}
  \label{eq:solms}
  \Delta m_{\odot }^2 \approx M_{\sss \nu_{\sss L}{\rm medium}}^2\approx 
\frac{W^4T^4\chi^2\xi^2\sVEV{\phi_{\sss WS}}^4}%
{4\sVEV{\phi_{\sss\rm B-L}}^2} \nn.
\end{equation}
The atmospheric mixing angle goes between the heaviest and the medium one:
\begin{eqnarray}
  \label{eq:atmms}
  \Delta m_{\rm atm}^2 &\approx& M_{\sss \nu_{\sss L}{\rm heavy}}^2-%
M_{\sss \nu_{\sss L}{\rm medium}}^2 \nonumber\\
&\approx& \frac{W^4T^4\chi^2\sVEV{\phi_{\sss WS}}^4}%
{4\sVEV{\phi_{\sss\rm B-L}}^2\xi^2}
\end{eqnarray}
From equations (\ref{eq:solms}) and (\ref{eq:atmms}) we find that the ratio
of solar and atmospheric mass square differences must be of the
order of $\xi^4$, say, about $10^{-4}$.

\subsection{Statistical calculation using random order unity factors}
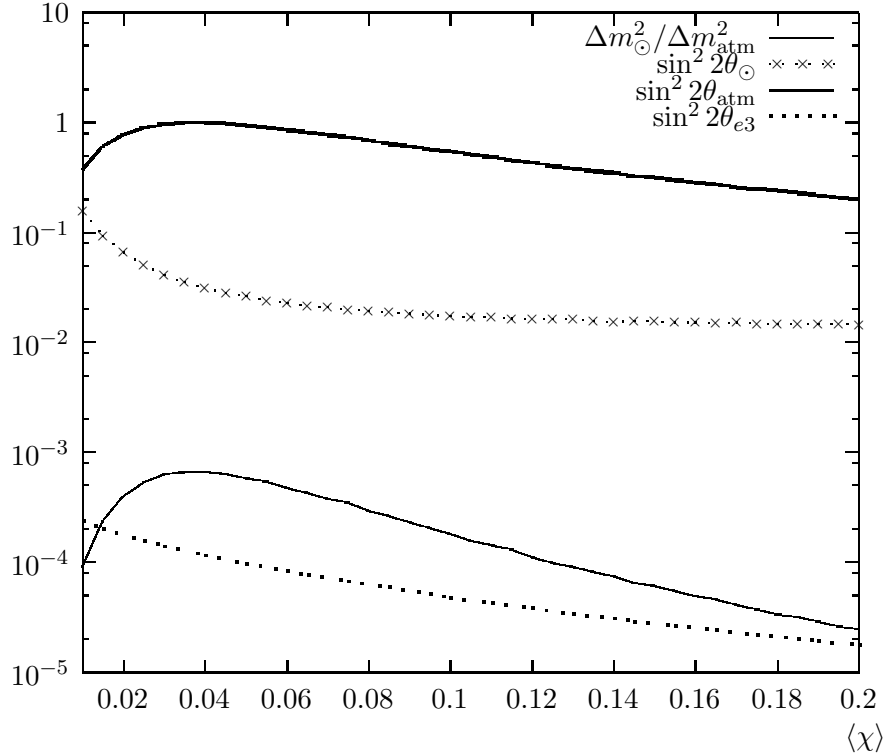
\begin{figure}[ht!]
\begin{center}
\label{fig:resuneu}
\setlength{\unitlength}{0.240900pt}
\ifx\plotpoint\undefined\newsavebox{\plotpoint}\fi
\sbox{\plotpoint}{\rule[-0.200pt]{0.400pt}{0.400pt}}%
\begin{picture}(1500,1200)(0,0)
\font\gnuplot=cmr10 at 10pt
\gnuplot
\sbox{\plotpoint}{\rule[-0.200pt]{0.400pt}{0.400pt}}%
\put(220.0,123.0){\rule[-0.200pt]{4.818pt}{0.400pt}}
\put(200,123){\makebox(0,0)[r]{$10^{-5}$}}
\put(1419.0,123.0){\rule[-0.200pt]{4.818pt}{0.400pt}}
\put(220.0,175.0){\rule[-0.200pt]{2.409pt}{0.400pt}}
\put(1429.0,175.0){\rule[-0.200pt]{2.409pt}{0.400pt}}
\put(220.0,244.0){\rule[-0.200pt]{2.409pt}{0.400pt}}
\put(1429.0,244.0){\rule[-0.200pt]{2.409pt}{0.400pt}}
\put(220.0,279.0){\rule[-0.200pt]{2.409pt}{0.400pt}}
\put(1429.0,279.0){\rule[-0.200pt]{2.409pt}{0.400pt}}
\put(220.0,296.0){\rule[-0.200pt]{4.818pt}{0.400pt}}
\put(200,296){\makebox(0,0)[r]{$10^{-4}$}}
\put(1419.0,296.0){\rule[-0.200pt]{4.818pt}{0.400pt}}
\put(220.0,348.0){\rule[-0.200pt]{2.409pt}{0.400pt}}
\put(1429.0,348.0){\rule[-0.200pt]{2.409pt}{0.400pt}}
\put(220.0,417.0){\rule[-0.200pt]{2.409pt}{0.400pt}}
\put(1429.0,417.0){\rule[-0.200pt]{2.409pt}{0.400pt}}
\put(220.0,452.0){\rule[-0.200pt]{2.409pt}{0.400pt}}
\put(1429.0,452.0){\rule[-0.200pt]{2.409pt}{0.400pt}}
\put(220.0,469.0){\rule[-0.200pt]{4.818pt}{0.400pt}}
\put(200,469){\makebox(0,0)[r]{$10^{-3}$}}
\put(1419.0,469.0){\rule[-0.200pt]{4.818pt}{0.400pt}}
\put(220.0,521.0){\rule[-0.200pt]{2.409pt}{0.400pt}}
\put(1429.0,521.0){\rule[-0.200pt]{2.409pt}{0.400pt}}
\put(220.0,589.0){\rule[-0.200pt]{2.409pt}{0.400pt}}
\put(1429.0,589.0){\rule[-0.200pt]{2.409pt}{0.400pt}}
\put(220.0,625.0){\rule[-0.200pt]{2.409pt}{0.400pt}}
\put(1429.0,625.0){\rule[-0.200pt]{2.409pt}{0.400pt}}
\put(220.0,642.0){\rule[-0.200pt]{4.818pt}{0.400pt}}
\put(200,642){\makebox(0,0)[r]{$10^{-2}$}}
\put(1419.0,642.0){\rule[-0.200pt]{4.818pt}{0.400pt}}
\put(220.0,694.0){\rule[-0.200pt]{2.409pt}{0.400pt}}
\put(1429.0,694.0){\rule[-0.200pt]{2.409pt}{0.400pt}}
\put(220.0,762.0){\rule[-0.200pt]{2.409pt}{0.400pt}}
\put(1429.0,762.0){\rule[-0.200pt]{2.409pt}{0.400pt}}
\put(220.0,798.0){\rule[-0.200pt]{2.409pt}{0.400pt}}
\put(1429.0,798.0){\rule[-0.200pt]{2.409pt}{0.400pt}}
\put(220.0,814.0){\rule[-0.200pt]{4.818pt}{0.400pt}}
\put(200,814){\makebox(0,0)[r]{$10^{-1}$}}
\put(1419.0,814.0){\rule[-0.200pt]{4.818pt}{0.400pt}}
\put(220.0,866.0){\rule[-0.200pt]{2.409pt}{0.400pt}}
\put(1429.0,866.0){\rule[-0.200pt]{2.409pt}{0.400pt}}
\put(220.0,935.0){\rule[-0.200pt]{2.409pt}{0.400pt}}
\put(1429.0,935.0){\rule[-0.200pt]{2.409pt}{0.400pt}}
\put(220.0,970.0){\rule[-0.200pt]{2.409pt}{0.400pt}}
\put(1429.0,970.0){\rule[-0.200pt]{2.409pt}{0.400pt}}
\put(220.0,987.0){\rule[-0.200pt]{4.818pt}{0.400pt}}
\put(200,987){\makebox(0,0)[r]{$1$}}
\put(1419.0,987.0){\rule[-0.200pt]{4.818pt}{0.400pt}}
\put(220.0,1039.0){\rule[-0.200pt]{2.409pt}{0.400pt}}
\put(1429.0,1039.0){\rule[-0.200pt]{2.409pt}{0.400pt}}
\put(220.0,1108.0){\rule[-0.200pt]{2.409pt}{0.400pt}}
\put(1429.0,1108.0){\rule[-0.200pt]{2.409pt}{0.400pt}}
\put(220.0,1143.0){\rule[-0.200pt]{2.409pt}{0.400pt}}
\put(1429.0,1143.0){\rule[-0.200pt]{2.409pt}{0.400pt}}
\put(220.0,1160.0){\rule[-0.200pt]{4.818pt}{0.400pt}}
\put(200,1160){\makebox(0,0)[r]{$10$}}
\put(1419.0,1160.0){\rule[-0.200pt]{4.818pt}{0.400pt}}
\put(284.0,123.0){\rule[-0.200pt]{0.400pt}{4.818pt}}
\put(284,82){\makebox(0,0){$0.02$}}
\put(284.0,1140.0){\rule[-0.200pt]{0.400pt}{4.818pt}}
\put(412.0,123.0){\rule[-0.200pt]{0.400pt}{4.818pt}}
\put(412,82){\makebox(0,0){$0.04$}}
\put(412.0,1140.0){\rule[-0.200pt]{0.400pt}{4.818pt}}
\put(541.0,123.0){\rule[-0.200pt]{0.400pt}{4.818pt}}
\put(541,82){\makebox(0,0){$0.06$}}
\put(541.0,1140.0){\rule[-0.200pt]{0.400pt}{4.818pt}}
\put(669.0,123.0){\rule[-0.200pt]{0.400pt}{4.818pt}}
\put(669,82){\makebox(0,0){$0.08$}}
\put(669.0,1140.0){\rule[-0.200pt]{0.400pt}{4.818pt}}
\put(797.0,123.0){\rule[-0.200pt]{0.400pt}{4.818pt}}
\put(797,82){\makebox(0,0){$0.1$}}
\put(797.0,1140.0){\rule[-0.200pt]{0.400pt}{4.818pt}}
\put(926.0,123.0){\rule[-0.200pt]{0.400pt}{4.818pt}}
\put(926,82){\makebox(0,0){$0.12$}}
\put(926.0,1140.0){\rule[-0.200pt]{0.400pt}{4.818pt}}
\put(1054.0,123.0){\rule[-0.200pt]{0.400pt}{4.818pt}}
\put(1054,82){\makebox(0,0){$0.14$}}
\put(1054.0,1140.0){\rule[-0.200pt]{0.400pt}{4.818pt}}
\put(1182.0,123.0){\rule[-0.200pt]{0.400pt}{4.818pt}}
\put(1182,82){\makebox(0,0){$0.16$}}
\put(1182.0,1140.0){\rule[-0.200pt]{0.400pt}{4.818pt}}
\put(1311.0,123.0){\rule[-0.200pt]{0.400pt}{4.818pt}}
\put(1311,82){\makebox(0,0){$0.18$}}
\put(1311.0,1140.0){\rule[-0.200pt]{0.400pt}{4.818pt}}
\put(1439.0,123.0){\rule[-0.200pt]{0.400pt}{4.818pt}}
\put(1439,82){\makebox(0,0){$0.2$}}
\put(1439.0,1140.0){\rule[-0.200pt]{0.400pt}{4.818pt}}
\put(220.0,123.0){\rule[-0.200pt]{293.657pt}{0.400pt}}
\put(1439.0,123.0){\rule[-0.200pt]{0.400pt}{249.813pt}}
\put(220.0,1160.0){\rule[-0.200pt]{293.657pt}{0.400pt}}
\put(1450,21){\makebox(0,0){$\sVEV{\chi}$}}
\put(220.0,123.0){\rule[-0.200pt]{0.400pt}{249.813pt}}
\put(1279,1120){\makebox(0,0)[r]{${\footnotesize\Delta m^2_{\odot}/\Delta m^2_{\rm atm}}$}}
\put(1299.0,1120.0){\rule[-0.200pt]{24.090pt}{0.400pt}}
\put(220,289){\usebox{\plotpoint}}
\multiput(220.58,289.00)(0.497,1.131){61}{\rule{0.120pt}{1.000pt}}
\multiput(219.17,289.00)(32.000,69.924){2}{\rule{0.400pt}{0.500pt}}
\multiput(252.58,361.00)(0.497,0.609){61}{\rule{0.120pt}{0.588pt}}
\multiput(251.17,361.00)(32.000,37.781){2}{\rule{0.400pt}{0.294pt}}
\multiput(284.00,400.58)(0.729,0.496){41}{\rule{0.682pt}{0.120pt}}
\multiput(284.00,399.17)(30.585,22.000){2}{\rule{0.341pt}{0.400pt}}
\multiput(316.00,422.58)(1.250,0.493){23}{\rule{1.085pt}{0.119pt}}
\multiput(316.00,421.17)(29.749,13.000){2}{\rule{0.542pt}{0.400pt}}
\multiput(348.00,435.61)(6.937,0.447){3}{\rule{4.367pt}{0.108pt}}
\multiput(348.00,434.17)(22.937,3.000){2}{\rule{2.183pt}{0.400pt}}
\multiput(412.00,436.95)(7.160,-0.447){3}{\rule{4.500pt}{0.108pt}}
\multiput(412.00,437.17)(23.660,-3.000){2}{\rule{2.250pt}{0.400pt}}
\multiput(445.00,433.93)(2.399,-0.485){11}{\rule{1.929pt}{0.117pt}}
\multiput(445.00,434.17)(27.997,-7.000){2}{\rule{0.964pt}{0.400pt}}
\multiput(477.00,426.93)(3.493,-0.477){7}{\rule{2.660pt}{0.115pt}}
\multiput(477.00,427.17)(26.479,-5.000){2}{\rule{1.330pt}{0.400pt}}
\multiput(509.00,421.92)(1.642,-0.491){17}{\rule{1.380pt}{0.118pt}}
\multiput(509.00,422.17)(29.136,-10.000){2}{\rule{0.690pt}{0.400pt}}
\multiput(541.00,411.93)(2.079,-0.488){13}{\rule{1.700pt}{0.117pt}}
\multiput(541.00,412.17)(28.472,-8.000){2}{\rule{0.850pt}{0.400pt}}
\multiput(573.00,403.93)(1.834,-0.489){15}{\rule{1.522pt}{0.118pt}}
\multiput(573.00,404.17)(28.841,-9.000){2}{\rule{0.761pt}{0.400pt}}
\multiput(605.00,394.93)(2.841,-0.482){9}{\rule{2.233pt}{0.116pt}}
\multiput(605.00,395.17)(27.365,-6.000){2}{\rule{1.117pt}{0.400pt}}
\multiput(637.00,388.92)(1.250,-0.493){23}{\rule{1.085pt}{0.119pt}}
\multiput(637.00,389.17)(29.749,-13.000){2}{\rule{0.542pt}{0.400pt}}
\multiput(669.00,375.93)(2.079,-0.488){13}{\rule{1.700pt}{0.117pt}}
\multiput(669.00,376.17)(28.472,-8.000){2}{\rule{0.850pt}{0.400pt}}
\multiput(701.00,367.92)(1.642,-0.491){17}{\rule{1.380pt}{0.118pt}}
\multiput(701.00,368.17)(29.136,-10.000){2}{\rule{0.690pt}{0.400pt}}
\multiput(733.00,357.93)(1.834,-0.489){15}{\rule{1.522pt}{0.118pt}}
\multiput(733.00,358.17)(28.841,-9.000){2}{\rule{0.761pt}{0.400pt}}
\multiput(765.00,348.93)(1.834,-0.489){15}{\rule{1.522pt}{0.118pt}}
\multiput(765.00,349.17)(28.841,-9.000){2}{\rule{0.761pt}{0.400pt}}
\multiput(797.00,339.92)(1.534,-0.492){19}{\rule{1.300pt}{0.118pt}}
\multiput(797.00,340.17)(30.302,-11.000){2}{\rule{0.650pt}{0.400pt}}
\multiput(830.00,328.93)(2.399,-0.485){11}{\rule{1.929pt}{0.117pt}}
\multiput(830.00,329.17)(27.997,-7.000){2}{\rule{0.964pt}{0.400pt}}
\multiput(862.00,321.93)(2.399,-0.485){11}{\rule{1.929pt}{0.117pt}}
\multiput(862.00,322.17)(27.997,-7.000){2}{\rule{0.964pt}{0.400pt}}
\multiput(894.00,314.92)(1.358,-0.492){21}{\rule{1.167pt}{0.119pt}}
\multiput(894.00,315.17)(29.579,-12.000){2}{\rule{0.583pt}{0.400pt}}
\multiput(926.00,302.93)(1.834,-0.489){15}{\rule{1.522pt}{0.118pt}}
\multiput(926.00,303.17)(28.841,-9.000){2}{\rule{0.761pt}{0.400pt}}
\multiput(958.00,293.93)(2.841,-0.482){9}{\rule{2.233pt}{0.116pt}}
\multiput(958.00,294.17)(27.365,-6.000){2}{\rule{1.117pt}{0.400pt}}
\multiput(990.00,287.93)(2.079,-0.488){13}{\rule{1.700pt}{0.117pt}}
\multiput(990.00,288.17)(28.472,-8.000){2}{\rule{0.850pt}{0.400pt}}
\multiput(1022.00,279.93)(2.399,-0.485){11}{\rule{1.929pt}{0.117pt}}
\multiput(1022.00,280.17)(27.997,-7.000){2}{\rule{0.964pt}{0.400pt}}
\multiput(1054.00,272.92)(1.642,-0.491){17}{\rule{1.380pt}{0.118pt}}
\multiput(1054.00,273.17)(29.136,-10.000){2}{\rule{0.690pt}{0.400pt}}
\multiput(1086.00,262.93)(3.493,-0.477){7}{\rule{2.660pt}{0.115pt}}
\multiput(1086.00,263.17)(26.479,-5.000){2}{\rule{1.330pt}{0.400pt}}
\multiput(1118.00,257.93)(2.079,-0.488){13}{\rule{1.700pt}{0.117pt}}
\multiput(1118.00,258.17)(28.472,-8.000){2}{\rule{0.850pt}{0.400pt}}
\multiput(1150.00,249.93)(2.079,-0.488){13}{\rule{1.700pt}{0.117pt}}
\multiput(1150.00,250.17)(28.472,-8.000){2}{\rule{0.850pt}{0.400pt}}
\multiput(1182.00,241.93)(3.493,-0.477){7}{\rule{2.660pt}{0.115pt}}
\multiput(1182.00,242.17)(26.479,-5.000){2}{\rule{1.330pt}{0.400pt}}
\multiput(1214.00,236.93)(1.893,-0.489){15}{\rule{1.567pt}{0.118pt}}
\multiput(1214.00,237.17)(29.748,-9.000){2}{\rule{0.783pt}{0.400pt}}
\multiput(1247.00,227.93)(2.399,-0.485){11}{\rule{1.929pt}{0.117pt}}
\multiput(1247.00,228.17)(27.997,-7.000){2}{\rule{0.964pt}{0.400pt}}
\multiput(1279.00,220.93)(2.079,-0.488){13}{\rule{1.700pt}{0.117pt}}
\multiput(1279.00,221.17)(28.472,-8.000){2}{\rule{0.850pt}{0.400pt}}
\multiput(1311.00,212.94)(4.575,-0.468){5}{\rule{3.300pt}{0.113pt}}
\multiput(1311.00,213.17)(25.151,-4.000){2}{\rule{1.650pt}{0.400pt}}
\multiput(1343.00,208.93)(2.399,-0.485){11}{\rule{1.929pt}{0.117pt}}
\multiput(1343.00,209.17)(27.997,-7.000){2}{\rule{0.964pt}{0.400pt}}
\multiput(1375.00,201.93)(2.079,-0.488){13}{\rule{1.700pt}{0.117pt}}
\multiput(1375.00,202.17)(28.472,-8.000){2}{\rule{0.850pt}{0.400pt}}
\multiput(1407.00,193.94)(4.575,-0.468){5}{\rule{3.300pt}{0.113pt}}
\multiput(1407.00,194.17)(25.151,-4.000){2}{\rule{1.650pt}{0.400pt}}
\put(380.0,438.0){\rule[-0.200pt]{7.709pt}{0.400pt}}
\put(1279,1079){\makebox(0,0)[r]{${\footnotesize\sin^22\theta_{\odot}}$}}
\multiput(1299,1079)(20.756,0.000){5}{\usebox{\plotpoint}}
\put(1399,1079){\usebox{\plotpoint}}
\put(220,849){\usebox{\plotpoint}}
\multiput(220,849)(12.966,-16.207){3}{\usebox{\plotpoint}}
\multiput(252,809)(16.109,-13.088){2}{\usebox{\plotpoint}}
\multiput(284,783)(17.601,-11.000){2}{\usebox{\plotpoint}}
\put(329.50,756.25){\usebox{\plotpoint}}
\multiput(348,747)(19.628,-6.747){2}{\usebox{\plotpoint}}
\multiput(380,736)(19.980,-5.619){2}{\usebox{\plotpoint}}
\put(427.57,723.23){\usebox{\plotpoint}}
\multiput(445,719)(20.507,-3.204){2}{\usebox{\plotpoint}}
\multiput(477,714)(20.276,-4.435){2}{\usebox{\plotpoint}}
\put(529.53,704.43){\usebox{\plotpoint}}
\multiput(541,703)(20.595,-2.574){2}{\usebox{\plotpoint}}
\put(591.43,697.85){\usebox{\plotpoint}}
\multiput(605,697)(20.595,-2.574){2}{\usebox{\plotpoint}}
\put(653.39,691.98){\usebox{\plotpoint}}
\multiput(669,691)(20.715,-1.295){2}{\usebox{\plotpoint}}
\put(715.53,688.09){\usebox{\plotpoint}}
\multiput(733,687)(20.715,-1.295){2}{\usebox{\plotpoint}}
\put(777.68,684.21){\usebox{\plotpoint}}
\multiput(797,683)(20.746,-0.629){2}{\usebox{\plotpoint}}
\multiput(830,682)(20.745,-0.648){2}{\usebox{\plotpoint}}
\put(881.35,679.79){\usebox{\plotpoint}}
\multiput(894,679)(20.745,-0.648){2}{\usebox{\plotpoint}}
\put(943.57,678.00){\usebox{\plotpoint}}
\multiput(958,678)(20.756,0.000){2}{\usebox{\plotpoint}}
\put(1005.77,676.52){\usebox{\plotpoint}}
\multiput(1022,675)(20.745,-0.648){2}{\usebox{\plotpoint}}
\put(1067.94,674.44){\usebox{\plotpoint}}
\multiput(1086,675)(20.756,0.000){2}{\usebox{\plotpoint}}
\put(1130.18,674.24){\usebox{\plotpoint}}
\multiput(1150,673)(20.756,0.000){2}{\usebox{\plotpoint}}
\multiput(1182,673)(20.745,-0.648){2}{\usebox{\plotpoint}}
\put(1233.89,672.60){\usebox{\plotpoint}}
\multiput(1247,673)(20.715,-1.295){2}{\usebox{\plotpoint}}
\put(1296.09,671.00){\usebox{\plotpoint}}
\multiput(1311,671)(20.745,-0.648){2}{\usebox{\plotpoint}}
\put(1358.34,670.00){\usebox{\plotpoint}}
\multiput(1375,670)(20.745,0.648){2}{\usebox{\plotpoint}}
\put(1420.57,670.15){\usebox{\plotpoint}}
\put(1439,669){\usebox{\plotpoint}}
\put(220,849){\makebox(0,0){${\sss \times}$}}
\put(252,809){\makebox(0,0){${\sss \times}$}}
\put(284,783){\makebox(0,0){${\sss \times}$}}
\put(316,763){\makebox(0,0){${\sss \times}$}}
\put(348,747){\makebox(0,0){${\sss \times}$}}
\put(380,736){\makebox(0,0){${\sss \times}$}}
\put(412,727){\makebox(0,0){${\sss \times}$}}
\put(445,719){\makebox(0,0){${\sss \times}$}}
\put(477,714){\makebox(0,0){${\sss \times}$}}
\put(509,707){\makebox(0,0){${\sss \times}$}}
\put(541,703){\makebox(0,0){${\sss \times}$}}
\put(573,699){\makebox(0,0){${\sss \times}$}}
\put(605,697){\makebox(0,0){${\sss \times}$}}
\put(637,693){\makebox(0,0){${\sss \times}$}}
\put(669,691){\makebox(0,0){${\sss \times}$}}
\put(701,689){\makebox(0,0){${\sss \times}$}}
\put(733,687){\makebox(0,0){${\sss \times}$}}
\put(765,685){\makebox(0,0){${\sss \times}$}}
\put(797,683){\makebox(0,0){${\sss \times}$}}
\put(830,682){\makebox(0,0){${\sss \times}$}}
\put(862,681){\makebox(0,0){${\sss \times}$}}
\put(894,679){\makebox(0,0){${\sss \times}$}}
\put(926,678){\makebox(0,0){${\sss \times}$}}
\put(958,678){\makebox(0,0){${\sss \times}$}}
\put(990,678){\makebox(0,0){${\sss \times}$}}
\put(1022,675){\makebox(0,0){${\sss \times}$}}
\put(1054,674){\makebox(0,0){${\sss \times}$}}
\put(1086,675){\makebox(0,0){${\sss \times}$}}
\put(1118,675){\makebox(0,0){${\sss \times}$}}
\put(1150,673){\makebox(0,0){${\sss \times}$}}
\put(1182,673){\makebox(0,0){${\sss \times}$}}
\put(1214,672){\makebox(0,0){${\sss \times}$}}
\put(1247,673){\makebox(0,0){${\sss \times}$}}
\put(1279,671){\makebox(0,0){${\sss \times}$}}
\put(1311,671){\makebox(0,0){${\sss \times}$}}
\put(1343,670){\makebox(0,0){${\sss \times}$}}
\put(1375,670){\makebox(0,0){${\sss \times}$}}
\put(1407,671){\makebox(0,0){${\sss \times}$}}
\put(1439,669){\makebox(0,0){${\sss \times}$}}
\put(1311,1079){\makebox(0,0){${\sss \times}$}}
\put(1351,1079){\makebox(0,0){${\sss \times}$}}
\put(1391,1079){\makebox(0,0){${\sss \times}$}}
\sbox{\plotpoint}{\rule[-0.400pt]{0.800pt}{0.800pt}}%
\put(1279,1038){\makebox(0,0)[r]{${\footnotesize\sin^22\theta_{\rm atm}}$}}
\put(1299.0,1038.0){\rule[-0.400pt]{24.090pt}{0.800pt}}
\put(220,913){\usebox{\plotpoint}}
\multiput(221.41,913.00)(0.503,0.577){57}{\rule{0.121pt}{1.125pt}}
\multiput(218.34,913.00)(32.000,34.665){2}{\rule{0.800pt}{0.563pt}}
\multiput(252.00,951.41)(0.902,0.506){29}{\rule{1.622pt}{0.122pt}}
\multiput(252.00,948.34)(28.633,18.000){2}{\rule{0.811pt}{0.800pt}}
\multiput(284.00,969.40)(1.536,0.512){15}{\rule{2.527pt}{0.123pt}}
\multiput(284.00,966.34)(26.755,11.000){2}{\rule{1.264pt}{0.800pt}}
\multiput(316.00,980.39)(3.365,0.536){5}{\rule{4.467pt}{0.129pt}}
\multiput(316.00,977.34)(22.729,6.000){2}{\rule{2.233pt}{0.800pt}}
\put(348,984.34){\rule{7.709pt}{0.800pt}}
\multiput(348.00,983.34)(16.000,2.000){2}{\rule{3.854pt}{0.800pt}}
\put(412,984.84){\rule{7.950pt}{0.800pt}}
\multiput(412.00,985.34)(16.500,-1.000){2}{\rule{3.975pt}{0.800pt}}
\put(445,982.84){\rule{7.709pt}{0.800pt}}
\multiput(445.00,984.34)(16.000,-3.000){2}{\rule{3.854pt}{0.800pt}}
\put(477,979.84){\rule{7.709pt}{0.800pt}}
\multiput(477.00,981.34)(16.000,-3.000){2}{\rule{3.854pt}{0.800pt}}
\put(509,976.34){\rule{6.600pt}{0.800pt}}
\multiput(509.00,978.34)(18.301,-4.000){2}{\rule{3.300pt}{0.800pt}}
\put(541,972.34){\rule{6.600pt}{0.800pt}}
\multiput(541.00,974.34)(18.301,-4.000){2}{\rule{3.300pt}{0.800pt}}
\put(573,968.34){\rule{6.600pt}{0.800pt}}
\multiput(573.00,970.34)(18.301,-4.000){2}{\rule{3.300pt}{0.800pt}}
\put(605,964.34){\rule{6.600pt}{0.800pt}}
\multiput(605.00,966.34)(18.301,-4.000){2}{\rule{3.300pt}{0.800pt}}
\multiput(637.00,962.06)(4.958,-0.560){3}{\rule{5.320pt}{0.135pt}}
\multiput(637.00,962.34)(20.958,-5.000){2}{\rule{2.660pt}{0.800pt}}
\multiput(669.00,957.06)(4.958,-0.560){3}{\rule{5.320pt}{0.135pt}}
\multiput(669.00,957.34)(20.958,-5.000){2}{\rule{2.660pt}{0.800pt}}
\put(701,950.34){\rule{6.600pt}{0.800pt}}
\multiput(701.00,952.34)(18.301,-4.000){2}{\rule{3.300pt}{0.800pt}}
\multiput(733.00,948.06)(4.958,-0.560){3}{\rule{5.320pt}{0.135pt}}
\multiput(733.00,948.34)(20.958,-5.000){2}{\rule{2.660pt}{0.800pt}}
\put(765,941.84){\rule{7.709pt}{0.800pt}}
\multiput(765.00,943.34)(16.000,-3.000){2}{\rule{3.854pt}{0.800pt}}
\multiput(797.00,940.06)(5.126,-0.560){3}{\rule{5.480pt}{0.135pt}}
\multiput(797.00,940.34)(21.626,-5.000){2}{\rule{2.740pt}{0.800pt}}
\put(830,933.34){\rule{6.600pt}{0.800pt}}
\multiput(830.00,935.34)(18.301,-4.000){2}{\rule{3.300pt}{0.800pt}}
\multiput(862.00,931.06)(4.958,-0.560){3}{\rule{5.320pt}{0.135pt}}
\multiput(862.00,931.34)(20.958,-5.000){2}{\rule{2.660pt}{0.800pt}}
\put(894,924.34){\rule{6.600pt}{0.800pt}}
\multiput(894.00,926.34)(18.301,-4.000){2}{\rule{3.300pt}{0.800pt}}
\multiput(926.00,922.06)(4.958,-0.560){3}{\rule{5.320pt}{0.135pt}}
\multiput(926.00,922.34)(20.958,-5.000){2}{\rule{2.660pt}{0.800pt}}
\put(958,915.34){\rule{6.600pt}{0.800pt}}
\multiput(958.00,917.34)(18.301,-4.000){2}{\rule{3.300pt}{0.800pt}}
\put(990,911.34){\rule{6.600pt}{0.800pt}}
\multiput(990.00,913.34)(18.301,-4.000){2}{\rule{3.300pt}{0.800pt}}
\put(1022,907.84){\rule{7.709pt}{0.800pt}}
\multiput(1022.00,909.34)(16.000,-3.000){2}{\rule{3.854pt}{0.800pt}}
\multiput(1054.00,906.06)(4.958,-0.560){3}{\rule{5.320pt}{0.135pt}}
\multiput(1054.00,906.34)(20.958,-5.000){2}{\rule{2.660pt}{0.800pt}}
\put(1086,900.34){\rule{7.709pt}{0.800pt}}
\multiput(1086.00,901.34)(16.000,-2.000){2}{\rule{3.854pt}{0.800pt}}
\put(1118,897.34){\rule{6.600pt}{0.800pt}}
\multiput(1118.00,899.34)(18.301,-4.000){2}{\rule{3.300pt}{0.800pt}}
\put(1150,893.34){\rule{6.600pt}{0.800pt}}
\multiput(1150.00,895.34)(18.301,-4.000){2}{\rule{3.300pt}{0.800pt}}
\put(1182,889.84){\rule{7.709pt}{0.800pt}}
\multiput(1182.00,891.34)(16.000,-3.000){2}{\rule{3.854pt}{0.800pt}}
\multiput(1214.00,888.06)(5.126,-0.560){3}{\rule{5.480pt}{0.135pt}}
\multiput(1214.00,888.34)(21.626,-5.000){2}{\rule{2.740pt}{0.800pt}}
\put(1247,882.34){\rule{7.709pt}{0.800pt}}
\multiput(1247.00,883.34)(16.000,-2.000){2}{\rule{3.854pt}{0.800pt}}
\put(1279,879.84){\rule{7.709pt}{0.800pt}}
\multiput(1279.00,881.34)(16.000,-3.000){2}{\rule{3.854pt}{0.800pt}}
\put(1311,876.84){\rule{7.709pt}{0.800pt}}
\multiput(1311.00,878.34)(16.000,-3.000){2}{\rule{3.854pt}{0.800pt}}
\put(1343,873.34){\rule{6.600pt}{0.800pt}}
\multiput(1343.00,875.34)(18.301,-4.000){2}{\rule{3.300pt}{0.800pt}}
\put(1375,869.34){\rule{6.600pt}{0.800pt}}
\multiput(1375.00,871.34)(18.301,-4.000){2}{\rule{3.300pt}{0.800pt}}
\put(1407,866.34){\rule{7.709pt}{0.800pt}}
\multiput(1407.00,867.34)(16.000,-2.000){2}{\rule{3.854pt}{0.800pt}}
\put(380.0,987.0){\rule[-0.400pt]{7.709pt}{0.800pt}}
\sbox{\plotpoint}{\rule[-0.500pt]{1.000pt}{1.000pt}}%
\put(1279,997){\makebox(0,0)[r]{${\footnotesize\sin^22\theta_{e3}}$}}
\multiput(1299,997)(20.756,0.000){5}{\usebox{\plotpoint}}
\put(1399,997){\usebox{\plotpoint}}
\put(220,361){\usebox{\plotpoint}}
\multiput(220,361)(19.229,-7.812){2}{\usebox{\plotpoint}}
\multiput(252,348)(19.811,-6.191){2}{\usebox{\plotpoint}}
\put(298.51,334.37){\usebox{\plotpoint}}
\multiput(316,330)(19.980,-5.619){2}{\usebox{\plotpoint}}
\multiput(348,321)(20.276,-4.435){2}{\usebox{\plotpoint}}
\put(399.29,309.78){\usebox{\plotpoint}}
\multiput(412,307)(20.304,-4.307){2}{\usebox{\plotpoint}}
\put(460.26,297.14){\usebox{\plotpoint}}
\multiput(477,294)(20.400,-3.825){2}{\usebox{\plotpoint}}
\put(521.46,285.66){\usebox{\plotpoint}}
\multiput(541,282)(20.400,-3.825){2}{\usebox{\plotpoint}}
\multiput(573,276)(20.400,-3.825){2}{\usebox{\plotpoint}}
\put(623.64,267.67){\usebox{\plotpoint}}
\multiput(637,266)(20.507,-3.204){2}{\usebox{\plotpoint}}
\put(685.28,258.96){\usebox{\plotpoint}}
\multiput(701,257)(20.276,-4.435){2}{\usebox{\plotpoint}}
\put(746.61,248.72){\usebox{\plotpoint}}
\multiput(765,247)(20.276,-4.435){2}{\usebox{\plotpoint}}
\multiput(797,240)(20.605,-2.498){2}{\usebox{\plotpoint}}
\put(849.16,233.60){\usebox{\plotpoint}}
\multiput(862,232)(20.595,-2.574){2}{\usebox{\plotpoint}}
\put(910.95,225.88){\usebox{\plotpoint}}
\multiput(926,224)(20.400,-3.825){2}{\usebox{\plotpoint}}
\put(972.48,216.64){\usebox{\plotpoint}}
\multiput(990,215)(20.595,-2.574){2}{\usebox{\plotpoint}}
\put(1034.36,209.84){\usebox{\plotpoint}}
\multiput(1054,208)(20.507,-3.204){2}{\usebox{\plotpoint}}
\multiput(1086,203)(20.595,-2.574){2}{\usebox{\plotpoint}}
\put(1137.33,197.19){\usebox{\plotpoint}}
\multiput(1150,196)(20.665,-1.937){2}{\usebox{\plotpoint}}
\put(1199.33,191.38){\usebox{\plotpoint}}
\multiput(1214,190)(20.521,-3.109){2}{\usebox{\plotpoint}}
\put(1261.09,183.68){\usebox{\plotpoint}}
\multiput(1279,182)(20.665,-1.937){2}{\usebox{\plotpoint}}
\put(1323.08,177.87){\usebox{\plotpoint}}
\multiput(1343,176)(20.595,-2.574){2}{\usebox{\plotpoint}}
\multiput(1375,172)(20.665,-1.937){2}{\usebox{\plotpoint}}
\put(1426.30,167.19){\usebox{\plotpoint}}
\put(1439,166){\usebox{\plotpoint}}
\end{picture}
\caption{The numerical results of the ratio of the 
    solar neutrino mass square difference to that 
    for the atmospheric neutrino oscillation, and  the 
    squared sine of the double of the solar neutrino 
    mixing angle, the atmospheric neutrino
    mixing angle and the mixing angle $\theta_{e3}$.}
  \end{center}
\end{figure}
\indent

In this subsection we will discuss the numerical 
calculation. The elements of the mass matrices are 
determined up to factors of order one to be a product of 
several Higgs VEVs measured in units of the fundamental scale 
$M_{\sss\rm Planck}$ - the Planck scale. 

We imagine that the mass matrix elements for, {\it e.g.}, the 
right-handed neutrino masses or the mass matrix $M_{\nu}^D$, 
are given by chain diagrams consisting of a backbone of 
fermion propagators for fermions with fundamental masses, 
with side ribs (branches) symbolising a Yukawa coupling to 
one of the Higgs field VEVs. 

We know neither the Yukawa couplings nor the precise 
masses of the fundamental mass fermions, but it is a basic 
assumption of the naturalness of our model that these 
couplings are of order unity and that the masses, also deviate 
from the 
Planck mass by factors of order unity. In the numerical 
evaluation of the consequences of the model we explicitly take 
into account these uncertain factors of order unity by 
providing each matrix element with an explicit random number 
$\lambda_{ij}$ - with a distribution so that its average 
$\sVEV{\log {\lambda_{ij}}}\approx 0$ and its spreading is
a factor two.

Then the calculation is performed with these numbers time
after time with different random number $\lambda_{ij}$-values 
and the results averaged in logarithms. A crude realisation of the 
distribution of these $\lambda_{ij}$ could be a flat 
distribution between $-2$ and $+2$, then provided also 
with a random phase (with flat distribution). 

Another ``detail'' is the use of a factor 
$\sqrt{{\#} {\rm diagrams}}$ multiplying 
the matrix elements, to take into account that, due to the 
possibility of permuting the Higgs field attachments in 
the chain-diagram, the number of different diagrams is 
roughly proportional to the number of such permutations 
\mbox{${\#}{\rm diagrams}$}. This is the correction introduced 
and studied for the charged mass matrices by C.D.~Froggatt, 
D.~Smith and one of us \cite{fnsnew}. In the philosophy of 
each diagram coming with a random order unity factor, the sum of 
\mbox{$\#{\rm diagrams}$} get of the order 
$\sqrt{{\#} {\rm diagrams}}$ bigger than a 
single diagram of that sort. But we counted these permutations
ignoring the field $S$. If we allowed both $S$ and 
$S^{\dagger}$ in the same diagram, the $\sqrt{\#{\rm diagrams}}$
factor could give arbitrarily large numbers. It turns out that 
these factors are especially important for some elements 
involving the electron-neutrino in the matrix 
$M_{\nu}^D$, which are suppressed by several factors, 
as then many permutations can be made.

Yet another detail is that the symmetric mass matrices
- occurring for the Majorana neutrinos - give rise to the same
off-diagonal term twice in the right-handed neutrino matrix in 
the effective Lagrangian, so we must multiply off-diagonal
elements with a factor $1/2$. But in the $M_{\nu}^D$-matrix,
columns and rows are related to completely 
different Weyl fields and of course a similar factor $1/2$ should not
be introduced.

Concerning the $\sqrt{\#{\rm diagrams}}$ factor for the
diagonal mass matrix terms in the symmetric matrices, 
{\it e.g.}~$M_R$, we shall remember that, contrary 
to what we shall do in non-symmetric matrices such as 
$M_{\nu}^D$, and the charged lepton ones. We must count 
diagrams with the Higgs fields attachment assigned in 
opposite order as only one diagram. The backbone in the 
diagram has no orientation and we shall count diagrams 
obtained from each other by inverting the sequence of 
the attached Higgs fields as only ONE diagram. Thus the 
diagonal elements will tend to have only half as many diagrams.

We give in Figure $1$ results obtained 
with $50,000$ random combinations averaged as a function 
of the small VEV $\sVEV{\chi}$ of the new Higgs field $\chi$.

In order to get the atmospheric mixing angle of the order 
of unity the range for $\sVEV{\chi}$ around the ``old'' 
Anti-GUT VEV $\sVEV{T}\approx0.07$ is suggested, so only 
this range is presented.

\section{Conclusion}
\label{sec:conclusion}

\indent
In this article we have made an extension of the Anti-GUT 
model to neutrinos by including see-saw 
$\nu_R$ at a scale of mass around $10^{12}~\GeV$. By this extension we 
introduced two more parameters, namely the vacuum expectation
values of two additional Higgs fields, $\phi_{\sss B-L}$ and
$\chi$. But from the neutrino oscillation data one extracts 
two mixing angles $\theta_{\odot}$ and $\theta_{\rm atm}$ and 
two mass square differences $\Delta m^2_{\odot}$ and 
${\Delta m^2_{\rm atm}}$, so in this sense we have two predictions:
\begin{eqnarray}
  \label{eq:ergebnis}
  \sin^22\theta_{\odot} &\approx& 3 \times 10^{-2}\\ 
  \frac{\Delta m^2_{\odot}}{\Delta m^2_{\rm atm}} &\approx& 6 \times 10^{-4}
\end{eqnarray}

These results are {\em only order of magnitude} estimates, and we shall 
count something like an uncertainty of $50\%$ for mixing angles
and masses and thus for the square, $\sin^22\theta$, $100\%$ and
{\it i.e.} a factor $2$ up or down and for the 
${\Delta m^2_{\odot}}/{\Delta m^2_{\rm atm}}$ , $\sqrt{2}\cdot 100\%$
meaning roughly a factor $3$ up or down,
\begin{eqnarray}
  \label{eq:ergebnis2}
    \sin^22\theta_{\odot} &=& (3 {\sss{ +3\atop -2 }})\times 10^{-2}\\ 
    \frac{\Delta m^2_{\odot}}{\Delta m^2_{\rm atm}} &=& (6 {\sss{ +11\atop -4 }})\times 10^{-4}\nn.
\end{eqnarray}%

These two small numbers both come from the parameter 
$\xi$ - the VEV in ``fundamental units'' of one of 
the $7$ Higgs fields in our model - which has 
already been fitted to the charged fermions in earlier 
works and which is essentially the Cabibbo angle measuring 
strange to up-quark weak transitions ($\xi\simeq0.1$ essentially 
$\sin\theta_{c}\simeq 0.22$). But it is also important 
for the success of our model that there has been room to 
put in the $\chi$ field, with which we could fix the 
atmospheric mixing angle to be of order unity (by taking 
$\chi\sim T$), as well as a parameter $\phi_{\sss B-L}$, 
the Higgs field VEV for breaking the gauged $B-L$ charge 
to fit the overall scale of observed neutrino masses. These 
factors in front of equations (\ref{eq:ergmitxi1}), (\ref{eq:ergmitxi2}) 
and (\ref{eq:ergmitxi3}) are results of 
our rather arbitrary averaging over random order unity factors and 
inclusion of diagram counting square root factors as put 
forward in reference \cite{fnsnew}. But in principle the 
factors in front are just of order unity:
\begin{eqnarray}
  \sin^22\theta_{\odot} &=& 3\,\xi^2\label{eq:ergmitxi1}\\
  \sin\theta_{c} &=& 1.8\,\xi\label{eq:ergmitxi2}\\
  \frac{\Delta m^2_{\odot}}{\Delta m^2_{\rm atm}} &=& 6\,\xi^4\label{eq:ergmitxi3}
\end{eqnarray}

\vspace*{-.3cm}
\begin{table}[!!b]
\caption{Number of parameters}
\label{Table2}
\begin{center}
\begin{tabular}{|c||ccc|c|c|} \hline
& \hspace{2mm}{\small ``Yukawa''}&\vline &{\small ``Neutrino''}\hspace{2mm} & {\small {\#} of parameters} & {\small {\#} of predictions}    \\ \hline
{\small Standard Model} & \hspace{2mm} $13$ & \vline & $4$ \hspace{2mm} & $17$  & ---\\ \hline
{\small ``Old'' Anti-GUT}   &  \hspace{2mm}  $4$ & \vline & ---$\sss{{^\ast}}$ \hspace{2mm} & $4^{\sss{\dagger}}$ & $9$ \\ \hline
{\small ``New'' Anti-GUT}   & & $6$ & & $6$ & $11$ \\ \hline
\end{tabular}
\end{center}
\vspace*{-0.4cm}
\begin{center}
{\small $^\ast$ The ``old''Anti-GUT cannot predict the neutrino oscillation.\\
$^{\dagger}$ Here we have not counted the neutrino oscillation parameters.}
\end{center}
\end{table}

We want to emphasise here that our model - extended Anti-GUT
as well as ``old'' Anti-GUT - is itself a good model in the 
sense that all coupling constants are order of unity 
except for Higgs fields VEVs, and thereby also Higgs masses 
giving rise to these VEVs. In the SM the most remarkable 
non-natural feature is the tremendously 
small Weinberg-Salam Higgs VEV compared to Planck or 
realistic GUT scales. If somehow we have to accept that 
there must be a mechanism in nature for making the Weinberg-Salam 
Higgs VEV very small, we also should admit that the other Higgs VEVs
could be very small. In our model we manage to interpret
the  second non-natural feature of most Yukawa couplings, namely, 
that they are very small to be also due to small Higgs field VEVs. 
In this way all small numbers come from VEVs in our model;
the rest is put to unity in Planck units. In this sense it is
``natural'' by the fact that it has only one source of small numbers, VEVs.
Even the gauge couplings are interpretable as being of order of unity,
if we follow our assumption, MPP \cite{mpp,mpp1}, which goes extremely 
well together with the present model.

It should, however, be stressed that to bring the neutrino 
oscillation into the model we have now two VEVs which are 
{\em extremely} small, namely not only the usual Weinberg-Salam Higgs 
VEV $\sVEV{\phi_{\sss WS}}\approx 173~\GeV$ but 
also $\sVEV{\phi_{\sss B-L}}\approx 10^{12}~\GeV$. In this 
sense the hierarchy problem is doubled. But this is not so 
much a special problem for our model; the problem is rather that 
neutrino oscillations point to a completely new scale model 
independently.

In the column ``Yukawa'' in Table $2$ we write the numbers 
of parameters with which the observed Yukawa couplings are 
fitted in the three models mentioned in first column, while 
``Neutrino'' stands for the number of the neutrino oscillation 
fitting parameters.

The column ``\# of parameters'' is the number of these two 
classes of parameters taken together. The last column 
``\# of predictions'' gives the number constraints predicted 
among measured parameters in SM and neutrino oscillations.

Including the old fits of the charge mass matrices we can say 
that we now fit, order-of-magnitude-wise, $17$ quantities 
($11$ observed  fermion masses or mass square differences, $5$ 
mixing angles and CP-violating phase of quarks) with $6$ 
parameters - the Higgs field VEVs. We can find from 
Table \ref{Table2} that we have used in the present article 
two parameters to fit four more quantities, thus 
gaining two predictions (the solar mixing angle and the ratio 
of the neutrino oscillation masses).

We should present here the order of magnitude of the right-handed 
neutrino masses and the mixing angle $\theta_{e3}$ 
(see in Figure $1$):
\begin{eqnarray}
  \label{eq:right-handedmixinge3}
  M_{R_{\nu_{\sss 1}}} &\approx& 10^{11} ~~{\rm\GeV}\nn,\\
  M_{R_{\nu_{\sss 2}}} &\approx& 10^{13} ~~{\rm\GeV}\nn, \\
  M_{R_{\nu_{\sss 3}}} &\approx& 10^{13}~~{\rm\GeV} \nn,\\
  \sin^22\theta_{e3}&\approx& 10^{-4}\nn.
\end{eqnarray}

Note that our model is very successful in describing neutrino 
oscillations and their mixing angles, but this model does not
have any good candidate for dark matter; the monopoles could be
such a candidate. We will study this problem in a forthcoming article.

\bigskip

\section*{Acknowledgements}

We would like to thank C.D.~Froggatt, M.~Gibson and S.~Lola for useful 
discussions. One of us (H.B.N.) wishes to thank 
W.~Buchm{\"u}ller for an important discussion of the see-saw mechanism. 
One of us (Y.T.) thanks the Theory Division of CERN for the 
hospitality extended to him during visits. H.B.N. wishes to 
thank the EU commission for grants SCI-0430-C (TSTS), CHRX-CT-94-0621, 
INTAS-RFBR-95-0567 and INTAS 93-3316(ext). Y.T. thanks 
the Danish Government for financial support.

\newpage 
%

%
\end{document}